\documentclass[aps,showpacs,twocolumn,nofootinbib,prd,superscriptaddress]{revtex4}

\usepackage{amsmath}
\usepackage{epsfig}
\usepackage[dvipsnames,usenames]{color}
\usepackage{hyperref}
\usepackage{breakurl}
\usepackage[normalem]{ulem}

\newcommand{\commentout}[1]{}

\newcommand{\MADM}{M_{\rm ADM}}
\newcommand{\JADM}{\vec{J}_{\rm ADM}}
\newcommand{\Mf}{M_{\rm f}}
\newcommand{\alpf}{\alpha_{\rm f}}
\newcommand{\atil}{\tilde{\alpha}}
\newcommand{\Rext}{r}

\newcommand{\sigQNM}{\sigma_{\ell m}}  % switched from \varpi
\newcommand{\sigtwotwoQNM}{\sigma_{22}} % switched from \varpi
\newcommand{\alpomtwotwo}{\Mf\alpf\sigtwotwoQNM}
\newcommand{\aw}{M\alpha\sigma} % switched from \varpi

\newcommand{\Ylm}[2]{\,_{0}Y_{#1}^{#2}}
\newcommand{\Ymtwolm}[2]{\,_{-2}Y_{#1}^{#2}}
\newcommand{\Smtwolm}[2]{\,_{-2}S_{#1}^{#2}}
\newcommand{\SYmtwolm}[2]{\,_{-2}\mathcal{Y}_{#1}^{#2}}

\newcommand{\beq}{\begin{equation}}
\newcommand{\eeq}{\end{equation}}
\newcommand{\ba}{\begin{array}}
\newcommand{\ea}{\end{array}}

\newcommand{\etal}{\emph{et al.\,}}

\begin{document}

\title{Decoding mode-mixing in black-hole merger ringdown}

\author{Bernard J. Kelly}
\affiliation{CRESST \& Gravitational Astrophysics Laboratory, NASA/GSFC, 8800 Greenbelt Rd., Greenbelt, MD 20771, USA}
\affiliation{Department of Physics, University of Maryland, Baltimore County, 1000 Hilltop Circle, Baltimore, MD 21250, USA}
\author{John G. Baker}
\affiliation{Gravitational Astrophysics Laboratory, NASA Goddard Space Flight Center, 8800 Greenbelt Rd., Greenbelt, MD 20771, USA}

\date{\today}

\begin{abstract}
Optimal extraction of information from gravitational-wave observations of binary black-hole
coalescences requires detailed knowledge of the waveforms. Current approaches for representing
waveform information are based on spin-weighted spherical harmonic decomposition. Higher-order
harmonic modes carrying a few percent of the total power output near merger can supply information
critical to determining intrinsic and extrinsic parameters of the binary. One obstacle to
constructing a full multi-mode template of merger waveforms is the apparently complicated behavior
of some of these modes; instead of settling down to a simple quasinormal frequency with decaying
amplitude, some $|m| \neq \ell$ modes show periodic bumps characteristic of mode-mixing. We analyze
the strongest of these modes -- the anomalous $(3,2)$ harmonic mode  -- measured in a set of binary
black-hole merger waveform simulations, and show that to leading order, they are due to a mismatch
between the \emph{spherical} harmonic basis used for extraction in 3D numerical relativity
simulations, and the \emph{spheroidal} harmonics adapted to the perturbation theory of Kerr black
holes. Other causes of mode-mixing arising from gauge ambiguities and physical properties of the
quasinormal ringdown modes are also considered and found to be small for the waveforms studied
here.
\end{abstract}

\pacs{
04.25.Dm, % numerical relativity
04.30.Db, % gravitational wave generation and sources
04.70.Bw, % classical black holes
95.30.Sf, % relativity and gravitation
97.60.Lf  % black holes (astrophysics)
}

\maketitle

\section{Introduction}
\label{sec:introduction}

Since the first successful simulation of black-hole binaries (BHBs) through late inspiral, merger,
and ringdown in 2005 \cite{Pretorius:2005gq,Campanelli:2005dd,Baker:2005vv}, theoretical interest
has centered on the resulting gravitational waveforms. A crucial tool in waveform studies has been
the analysis of the radiation wave pattern in spherical harmonic components. This decomposition is
useful both in the physical interpretation of the radiation, and in structuring the waveform
information content for the development of approximate analytic or empirical encodings.

The self-consistency of results for the dominant quadrupole waveforms across numerical codes was
quickly established \cite{Baker:2007fb,Hannam:2009hh}, enabling rapid study of the basic
characteristics of mergers
\cite{Campanelli:2006uy,Buonanno:2006ui,Boyle:2007ft,Scheel:2008rj,Gonzalez:2008bi,Lousto:2010tb,Centrella:2010mx,Hinder:2010vn}
Researchers soon began to build analytic template models compatible with these numerical results as
well as with the post-Newtonian (PN) at earlier times, to provide relatively quick waveforms for
specified BHB source masses and spins \cite{Ajith:2009bn,Pan:2009wj,Santamaria:2010yb}. While
expected to be sufficient for \emph{detection} of BHB mergers, quadrupole-only templates will not
lock down most of the intrinsic (masses, spin magnitudes \& directions) and extrinsic (sky
position, phase) BHB system parameters. To gain an understanding of these parameters requires a
richer template bank, one that includes all of the relevant angular modes of the signal
\cite{Arun:2007qv,Arun:2007hu,Trias:2008pu,McWilliams:2009bg}.

\begin{figure}
\includegraphics*[width=3.5in]{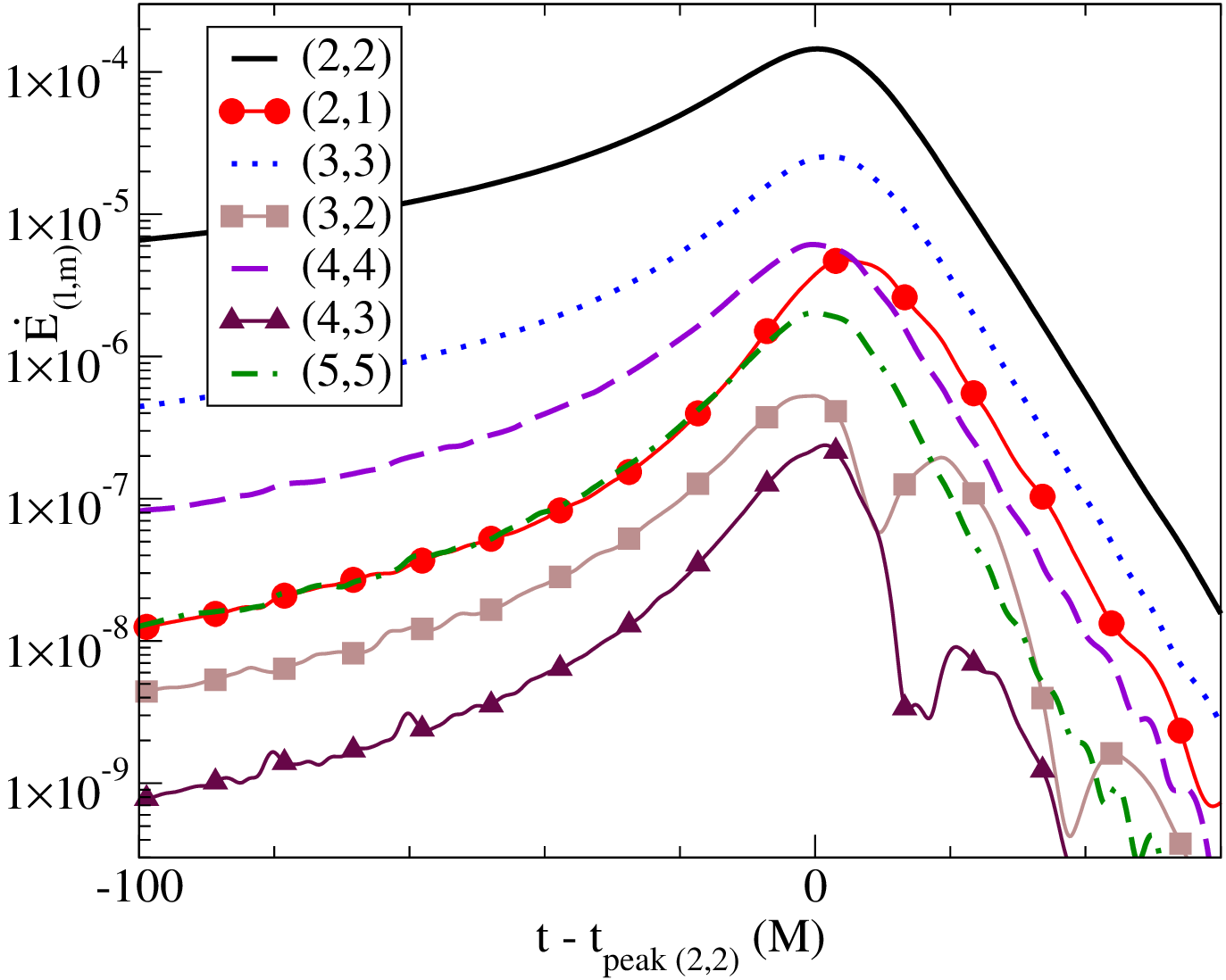}
\caption{Gravitational-wave luminosity from the merger and ringdown of a nonspinning
black-hole binary of mass ratio 4:1, decomposed by harmonic mode.}
\label{fig:relative_power_X4}
\end{figure}

Working with a spherical harmonic basis of spin-weight $s=-2$ \cite{Goldberg:1966uu,Wiaux:2005fm},
several studies \cite{Berti:2007fi,Berti:2007nw,Baker:2008mj,Kelly:2011bp,Pan:2011gk} have found
that after the dominant quadrupole $(\ell=2,m=\pm2)$ modes, the next most important modes tend to
be the higher $m=\pm\ell$ modes: $(3,\pm3)$, $(4,\pm4)$, etc., though odd-$m$ modes are sometimes
suppressed by symmetry. We have also seen, however, that certain $m < |\ell|$ modes can be
important. Prominent amongst these are the $(2,\pm1)$ and $(3,\pm2)$ modes.
Figure~\ref{fig:relative_power_X4} shows the radiative power for the most important modes in the
case of the merger of a 4:1 nonspinning BHB. Here we see that the $(2,1)$ mode has actually
overtaken the $(5,5)$ mode in importance by merger time.

A key feature of BHB mergers exposed through the spherical harmonic decomposition waveform studies
is the rather clean separation of the sometimes complicated mix of signal frequencies, achieved by
angular-mode decomposition. Even when typical observers would measure complicated wave shapes
combining several frequency harmonics, these harmonics largely reduce to slowly evolving sinusoids
in each spherical harmonic component mode. To a very good approximation, this structure holds
consistently through the inspiral, merger, and ringdown
\cite{Baker:2001sf,Baker:2002qf,Berti:2007fi,Baker:2006yw}.
This pattern of \emph{frequency separation} is extremely convenient in allowing relatively simple
encodings of the waveform information in analytic models.

Partly because of these properties, angular-mode decomposition has become a standard approach to
comparing waveform simulations with each other, with analytic post-Newtonian calculations, and with
developing empirical waveform template models. These uses of the decomposition technique have
elevated its significance from its beginning as an interpretive convenience to its current status
as an essential component of how we quantitatively understand gravitational-wave signals. Thus we
must be aware of the possibility that artifacts of arbitrary choices in the details of the
decomposition procedure may interfere with our quantitative understanding of the waveforms
themselves.

Such concerns are particularly notable when we see unusual features in the decomposed waveforms
seeming to violate the \emph{a posteriori} expectation of clean separation of frequencies. Several
authors \cite{Buonanno:2006ui,Schnittman:2007ij,Baker:2008mj,Pan:2011gk,Kelly:2011bp} have noted
that the $(3,2)$ mode in particular typically seems to break from this simple pattern, showing
unusual post-merger features that require investigation and resolution before a useful model can be
developed. In some of the earliest merger simulations, Buonanno \etal \cite{Buonanno:2006ui}
already noted the presence in the post-merger ``ringdown'' $(3,2)$ mode of both $(3,2)$ \emph{and}
$(2,2)$ quasinormal-mode (QNM) frequencies.

Existing multiple-mode template banks for low-eccentricity coalescences generally assume a
monotonic increase in frequency, and a simple single-peaked corresponding amplitude for each mode.
Although the $(3,2)$ mode is generically much weaker than the first few $\ell=m$ modes, if such
template models are applied to it na\"ively, they may suffer significant biases in their fitting
parameters. How serious the effect might be on parameter-estimation studies using these template
banks is unknown at the time of writing.

In this paper, we investigate these $(3,2)$-mode anomalies, with a survey of 3D numerical
simulations of the merger of various comparable-mass BHBs with non-precessing spins, exploring a
range of possible ``causes''. We find that the dominant part of the measured mode-mixing that
underlies the anomalous effect can be attributed to our use of spherical harmonics rather than the
\emph{spheroidal} harmonics expected by Teukolsky perturbation theory.

The remainder of this paper is laid out as follows: In Sec.~\ref{sec:bump_measured}, we review
the numerical evidence for mode-mixing in existing $(3,2)$ evolutions, and show how well it is
captured by a simple two-mode phenomenological model for the ringdown waveform segment. In
Sec.~\ref{sec:mode_mix_model}, we discuss general models for why mode-mixing should be expected,
including effects of coordinate distortions in the radiation extraction spheres, and of
ill-adapted harmonic basis functions in the radiation decomposition. In
Sec.~\ref{sec:simulations}, we introduce our set of expanded numerical evolutions, arranged into
``equivalence classes'' of common end-state Kerr spins, which we analyze in
Sec.~\ref{sec:analysis}, fitting the measured contributions of two-mode models to our models. We
conclude in Sec.~\ref{sec:discuss} with discussion on the application of these results to more
general late-merger-ringdown models, such as the \emph{implicit rotating source} model of
Refs.~\cite{Baker:2008mj,Kelly:2011bp}. We present a detailed description of our selection of
equivalence classes of binaries in Appendix~\ref{sec:equiv_ID}.

%%%%%%%%%%%%%%%%%%%%%%%%%%%%%%%%%%%%%%%%%%%%%%%%%%%%%%%%%%%%%%%%%%%%%%%%%%%%%%%%%%%%%%%%%%%%%%%%%%%
%%%%%%%%%%%%%%%%%%%%%%%%%%%%%%%%%%%%%%%%%%%%%%%%%%%%%%%%%%%%%%%%%%%%%%%%%%%%%%%%%%%%%%%%%%%%%%%%%%%
%%%%%%%%%%%%%%%%%%%%%%%%%%%%%%%%%%%%%%%%%%%%%%%%%%%%%%%%%%%%%%%%%%%%%%%%%%%%%%%%%%%%%%%%%%%%%%%%%%%

\section{Bumps in Numerical $(3,2)$ Modes}
\label{sec:bump_measured}

The first gravitational waveforms extracted from numerical simulations were the dominant $(2,\pm2)$
modes, whose early-inspiral behavior was expected to match the quadrupole radiation predicted by
quasi-Newtonian and post-Newtonian theory. Once these had been shown to be robust and universal
across codes \cite{Baker:2007fb,Hannam:2009hh}, some groups turned their attention to the
subdominant modes. Analyzing the subdominant modes of equal-mass binaries, Buonanno \etal
\cite{Buonanno:2006ui} reported that an accurate fit of the $(3,2)$ mode for the ringdown stage of
effective-one-body (EOB) waveforms requires the addition of the fundamental $(2,2)$ quasinormal
frequency. When Baker \etal \cite{Baker:2008mj} looked at a set of mergers of nonspinning
black-hole binaries with mass ratios in the range 1:1 to 6:1, they noted that one of the leading
subdominant modes, $(3,2)$, showed an unusual bumpiness just after merger over a range of parameter
space. This bumpiness manifested in both the frequency and amplitude, and appeared to persist with
both increased resolution and extraction radius, thus constituting a robust pattern of excursions
from the frequency separation dominating the $\ell=m$ modes. More recent work by Kelly \etal
\cite{Kelly:2011bp} shows the same anomaly in equal-mass binaries with non-precessing spins (i.e.,
the spins are aligned/anti-aligned with the orbital angular momentum).

\begin{figure}
\includegraphics*[width=3.5in]{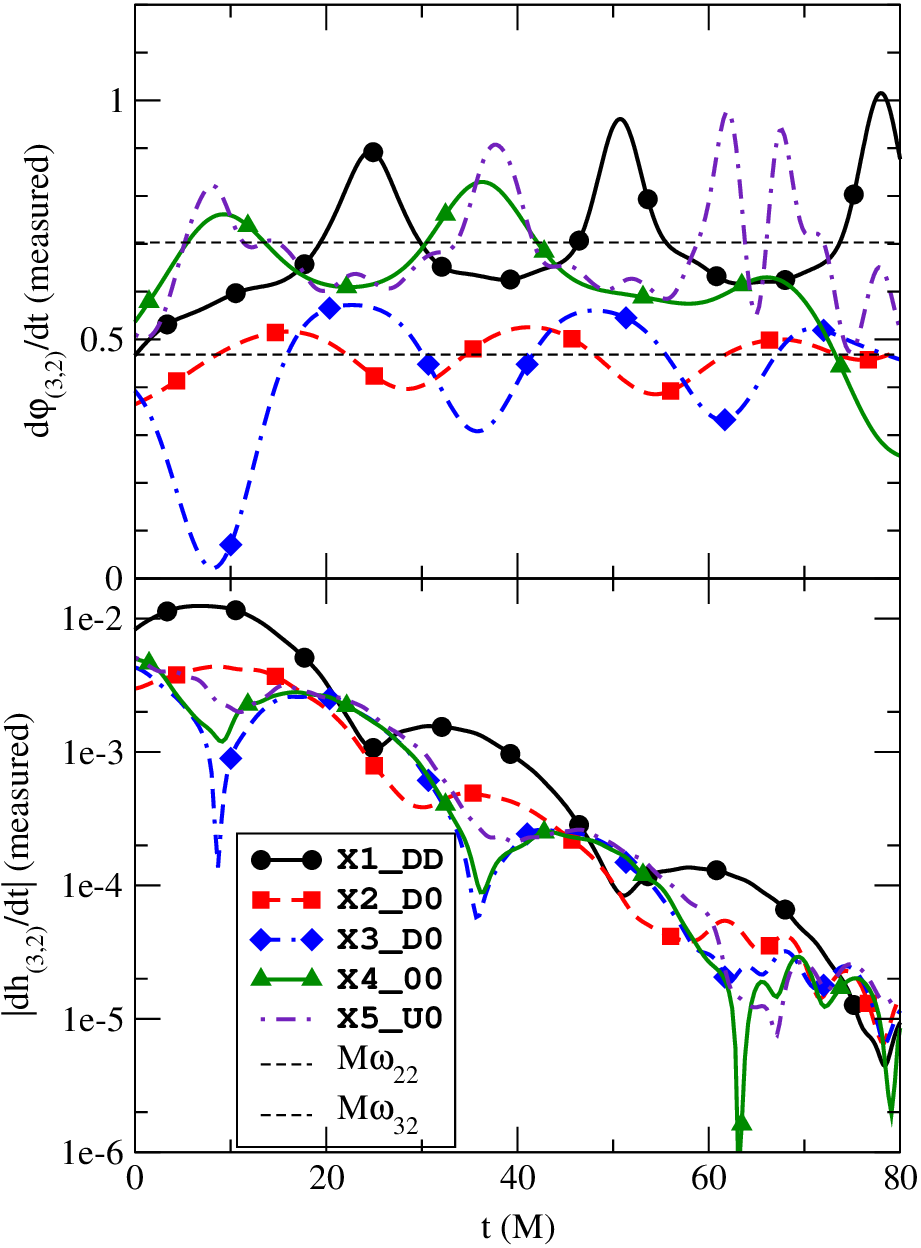}
\caption{Post-merger frequency [top panel] and amplitude [bottom panel] of the
numerically measured $(3,2)$ mode for a set of ``4:1-equivalent'' evolutions,
resulting in a final black hole with dimensionless spin $\alpf \approx 0.475$,
matching that of a 4:1 nonspinning binary merger. The data sets have been
shifted in time so that $t=0$ corresponds to peak amplitude of the dominant
$(2,2)$ mode. The two dashed (black) horizontal lines in the top panel mark the
fundamental QNM frequencies $\omega_{22}$ (lower) and $\omega_{32}$ (higher)
for a Kerr hole of the same final spin.}
\label{fig:om_amp32_data_X4_00}
\end{figure}

Examples of these more complicated waveform features are shown in
Fig.~\ref{fig:om_amp32_data_X4_00}, where we plot waveform frequency (top panel) and amplitude
(bottom panel) of the measured $(3,2)$ mode for the merger of a nonspinning 4:1 binary, as well as
for the mergers of several other BH configurations with the same final dimensionless spin
($\alpf \approx 0.475$). We also mark the expected real QNM $(2,2)$ and $(3,2)$ frequencies,
$\omega_{22}$ and $\omega_{32}$ for a Kerr black hole of this spin. From the time of peak amplitude
($t=0$ here) until the waveforms start to degrade around $60 M$ later, the frequency seems to
oscillate around one or other of these two QNM frequencies, rather than locking onto the higher
$\omega_{32}$, as for other modes. These oscillations appear in the strain $h$ and its
time-derivatives; we choose to study strain-rate, $\dot{h}(t)$, waveforms, which we decompose into
modes $\dot{h}_{(\ell,m)}(t)=|\dot{h}_{(\ell,m)}| \exp(i\varphi_{(\ell,m)})$, with instantaneous
frequencies $\dot\varphi_{(\ell,m)}$.

We can model the more complicated ringdown waveform features by expressing the $(2,2)$ mode as a
pure QNM ringdown, and the measured $(3,2)$ mode as a linear combination of QNM ringdowns:
\begin{align}
\dot{h}_{(2,2)}^{\rm model} &= A_{22} e^{i(\sigma_{22} t + \delta_{22})}, \label{eq:22mod_QNM_model}\\
\dot{h}_{(3,2)}^{\rm model} &= A_{32} e^{i(\sigma_{32} t + \delta_{32})} + \rho_{32} A_{22} e^{i(\sigma_{22} t + \delta_{22})}. \label{eq:32mod_QNM_model}
\end{align}
Here $\sigQNM \equiv \omega_{\ell m} + i/\tau_{\ell m}$ is the full complex QNM frequency, and
$\rho_{32} \equiv \rho_0 \exp (i \zeta)$ is a constant complex-valued parameter indicating the
mixing of the $(2,2)$ QNM mode into the measured $(3,2)$ mode. The modeled $(3,2)$ mode frequency
and amplitude are then:
\begin{align}
\dot\varphi_{(3,2)}^{\rm model}(t)            =& \omega_{32} + \frac{\varepsilon(t)^2 \Delta_R}{F(t)} \label{eq:freq_mod_theory}\\
                                               &- \frac{\varepsilon(t) \left[ \Delta_R \cos(\Delta_R t + \delta) + \Delta_I \sin(\Delta_R t + \delta) \right]}{F(t)}, \nonumber \\
\left| \dot{h}_{(3,2)}^{\rm model}(t) \right| =& A_{32} e^{-t/\tau_{32}} \sqrt{F(t)}. \label{eq:amp_mod_theory}
\end{align}
where $F(t) \equiv 1 + 2 \varepsilon(t) \cos(\Delta_R t + \delta) + \varepsilon(t)^2$,
$\Delta_R \equiv \omega_{32} - \omega_{22}$, $\Delta_I = 1/\tau_{32} - 1/\tau_{22}$,
$\varepsilon(t) \equiv \rho_0 A_{22}/A_{32} \exp(\Delta_I t) \equiv \varepsilon_0 \exp(\Delta_I t)$,
and $\delta \equiv \delta_{32} - \delta_{22} - \zeta$.

For a given mass and spin, the QNM frequencies, $\omega_{22}$ and $\omega_{32}$, and the damping
times, $\tau_{22}$ and $\tau_{32}$, are values known from black-hole perturbation theory.
Typically, $\tau_{22}\approx \tau_{32}$, so that $\Delta_I$ is somewhat smaller than $\Delta_R$,
allowing a beat-like effect to persist over several cycles. Fixing these leaves just two free
parameters for the frequency: $\varepsilon_0 \equiv \rho_0 A_{22}/A_{32}$, the initial ratio of
contributing amplitudes, and $\delta$, the initial phase difference, as well as one more amplitude
parameter, $A_{32}$.

\begin{figure}
\includegraphics*[width=3.5in]{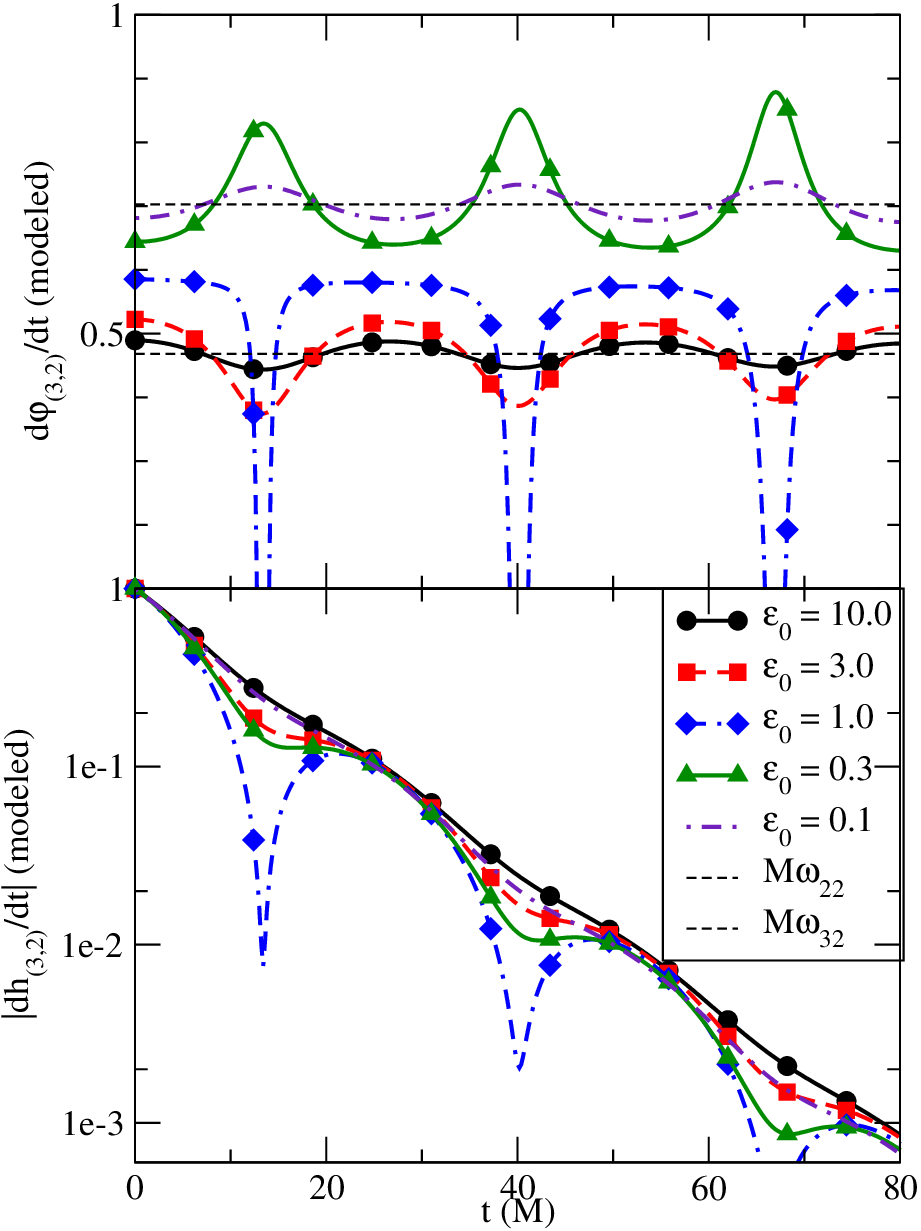}
\caption{Possible shapes from Eqs. \eqref{eq:freq_mod_theory}-\eqref{eq:amp_mod_theory}
for the modeled $(3,2)$
mode frequency [top panel] and amplitude [bottom panel] resulting from a merger
with an $\alpf = 0.475$ endpoint. All curves assume zero phase difference
$\delta$, and overall amplitude is arbitrarily scaled to unity at $t=0$. As
with Fig.~\ref{fig:om_amp32_data_X4_00}, the two dashed (black) horizontal
lines in the top panel mark the fundamental QNM frequencies $\omega_{22}$
(lower) and $\omega_{32}$ (higher) for a Kerr hole of the same final spin.}
\label{fig:om_amp32_theory}
\end{figure}

Evidently, the characteristic shape of the modeled (3,2) mode frequency plots will depend on the
relative magnitude of the modal contributions: for $\varepsilon_0 \ll 1$, the frequency will
oscillate (approximately) sinusoidally about $\omega_{32}$; for $\varepsilon_0 \gg 1$, the
oscillation will be about $\omega_{22}$; for intermediate values, the oscillatory shape will be
more complex. In the left panel of Fig.~\ref{fig:om_amp32_theory}, we demonstrate these shapes for
a Kerr hole of spin $\alpf = 0.475$, the same 4:1 end-state spin as in
Fig.~\ref{fig:om_amp32_data_X4_00}. Similarly, the right panel shows the corresponding modal
amplitude shape for the same end-state hole. Again, the most extreme bumps in amplitude occur when
the $(2,2)$ and $(3,2)$ modes have comparable amplitude contributions ($\varepsilon_0 \sim 1$).
These theoretical curves should be compared with the numerically measured mixing in
Fig.~\ref{fig:om_amp32_data_X4_00}.

%%%%%%%%%%%%%%%%%%%%%%%%%%%%%%%%%%%%%%%%%%%%%%%%%%%%%%%%%%%%%%%%%%%%%%%%%%%%%%%%%%%%%%%%%%%%%%%%%%%
%%%%%%%%%%%%%%%%%%%%%%%%%%%%%%%%%%%%%%%%%%%%%%%%%%%%%%%%%%%%%%%%%%%%%%%%%%%%%%%%%%%%%%%%%%%%%%%%%%%
%%%%%%%%%%%%%%%%%%%%%%%%%%%%%%%%%%%%%%%%%%%%%%%%%%%%%%%%%%%%%%%%%%%%%%%%%%%%%%%%%%%%%%%%%%%%%%%%%%%

\section{Possible Causes of Mode-Mixing}
\label{sec:mode_mix_model}

The bumpy features seen in the measured $(3,2)$ mode are a clear exception to the general rule that
each angular mode encodes sinusoidal waves with just one slowly evolving frequency component, the
phenomenon we refer to as \emph{frequency separation}. In Fig.~\ref{fig:om_amp32_theory}, we showed
that a combination of the fundamental $(3,2)$ and $(2,2)$ quasinormal-mode frequencies produces
similar features. More generally, there are indications that such mixing occurs among other modes,
especially other higher-order $m=2$ modes, which likewise seem prone to coupling to the dominant
mode. Here we ask the basic question: is this mode-mixing a fundamental property of the radiation,
or some kind of an artifact, and if so, what kind? 

We consider various hypotheses to explain this mode-mixing effect violating our empirical frequency
separation rule. The first, which we label \emph{physical mixing}, is simply that the frequency
separation rule does not physically hold to sufficiently high precision; that is we are are perhaps
seeing a nonlinear effect in the radiation-generation process underlying the $(3,2)$ mode. Under
this assumption, no choice of fixed or slowly evolving angular basis could be expected to yield the
kind of frequency separation we see in other cases. Near the merger where nonlinear physics is
dominant, it is difficult to make any strong argument for expecting frequency separation. Indeed,
we would be surprised to \emph{not} find violations of this assumption as we probe beyond the first
few orders of magnitude in waveform precision.

In the linear ringdown dynamics where this investigation is focused, some degree of physical
frequency separation can be expected, based on the separability of the Teukolsky equation, which
describes small distortions of a stationary black-hole spacetime. The scale of physical linear
mode-mixing can be quantified by careful consideration of quasinormal modes.

The alternative hypothesis is that the mixing is an artifact of our analysis, arising from choices
that we make in setting up the angular-mode decomposition. Perhaps our basis is not quite optimal,
but we can find some other basis in which we more precisely recover frequency separation. Indeed,
given the freedom available in selecting such a representation, we have little grounds for
supposing that our first guess would be optimal. Here we consider two classes of choices in how to
represent the space of gravitational radiation waveforms, which, in the full sense, has angular and
retarded-time dimensions.

The first choice we make is in how we define the spheres on which angular harmonic decomposition
will be conducted. Within the structure of asymptotically flat spacetimes, gauge freedom in the
choice of constant-retarded-time spheres can yield a frequency-dependent mode-mixing effect in the
decomposed waveforms. This ambiguity arises from the freedom to re-parameterize the proper-time
coordinate, the so-called ``supertranslations'' subgroup of the Bondi-Metzner-Sachs gauge group for
outgoing radiation. We describe this possibility of \emph{supertranslation gauge mixing} in more
detail below. We may generally expect that mode-mixing of this sort will be most evident in the
late merger, where wavelengths are shortest.

The next choice we make is in choosing the family of angular basis functions on the extraction
spheres. In this case, the mixing arises if our chosen family of modal basis functions used for
radiation extraction differs from the optimal one in which frequency separation is best
approximated. It is common to apply a spin-weighted spherical-harmonic basis, but a different
choice may be motivated for the ringdown signals. Indeed, the separation of the Teukolsky equation
is not achieved in a spin-weighted spherical harmonic basis, but in a spin-weighted
\emph{spheroidal}-harmonic basis. It has been suggested \cite{Buonanno:2006ui,Berti:2007fi} that
this difference explains the sort of waveform phenomena we consider, though this has not been
demonstrated. We label this effect \emph{angular-basis mixing}.  

In the next subsections, we consider these possible mixing effects in detail, preparing for a
quantitative study of the evidence for these effects in numerical data in
Sec.~\ref{sec:analysis}.

\subsection{Gauge Effects}
\label{ssec:gauge_mixing}

To understand the effect we are calling \emph{supertranslation gauge mixing}, we must make a brief
detour to describe the gauge freedom in the representation of an outgoing radiation field
approaching future null infinity in an asymptotically flat spacetime. Consider such a spacetime in
standard retarded-time coordinates $\{u, r, \theta,\phi\}$.  Scaled by $r$, the outgoing radiation
field propagates outward on null rays labeled by $u,\theta$, and $\phi$. Each polarization
component can thus be described by a function of these variables. The Bondi-Metzner-Sachs (BMS)
\cite{Bondi:1962px,Sachs:1962wk} group describes gauge transformations among these variables of the
form
\beq
\theta' = \theta'(\theta,\phi), \;\; \phi' = \phi'(\theta,\phi),\;\; u' = K(\theta,\phi)\left(u-\alpha(\theta,\phi)\right),\nonumber
\eeq
where $(\theta,\phi)\rightarrow(\theta',\phi')$ is a conformal transformation on a constant-$u$
sphere with conformal factor $K$. 

For concreteness in the context of numerical relativity simulations, we note that it is common to
make these gauge choices by specifying an ``extraction sphere'' located sufficiently far from the
source where radiation field calculations are realized. The effect of one class of BMS
transformations, amounting to rotations of the extraction sphere, has been identified as an
important concern when the choice of axis is not fixed by symmetry
\cite{Gualtieri:2008ux,Campanelli:2008nk,Schmidt:2010it,OShaughnessy:2011fx,OShaughnessy:2012vm,Schmidt:2012rh}.
However, the simulations in this study involve nonprecessing mergers, with no ambiguity in defining
the orientation of the extraction sphere.

But what happens if we make a small radial perturbation of the extraction sphere? It is clear that
sufficiently small distortions of larger extraction spheres would have  negligible impact on the
intrinsic geometry of the sphere. The gauge effects of such distortions are described by a subset
of the BMS transformations, known as \emph{supertranslations}, with $\theta'=\theta, \phi'=\phi$,
and $K=1$.

Now consider the effect of a supertranslation on a gravitational waveform $\psi(u,\theta,\phi)$.
Here we will make the additional assumption that $\alpha(\theta,\phi)$ is sufficiently small that
we can approximate the effect of the supertranslation by 
\beq
\label{eq:supertran}
\psi(u',\theta,\phi) \approx \psi(u,\theta,\phi) + \alpha(\theta,\phi) \frac{\partial}{\partial u}\psi(u,\theta,\phi),
\eeq
and we can expand the supertranslation in terms of (scalar) spherical harmonics:
\beq
\alpha(\theta,\phi) = \sum_{LM} b_{LM}\,{\,} \Ylm{L}{M}(\theta,\phi).
\eeq
Then from \eqref{eq:supertran}, the measured radiation modes will be perturbed as follows:
\beq
\psi_{\ell m}(u') \approx \psi_{\ell m}(u) + \sum_{\ell'm'}{C_{\ell m\ell'm'}\frac{\partial}{\partial u}\psi_{\ell'm'}(u)},
\label{eq:mode_genperturb_BMS}
\eeq
where
\begin{align}
C_{\ell m\ell'm'} = \sum_{L M} & b_{LM} \oint {\,}\Ylm{L}{M} {\,}\Ymtwolm{\ell'}{m'} {\,}\Ymtwolm{\ell}{m*} {\,} d\Omega \nonumber \\
                  = \sum_{L M} & b_{LM} \left[\frac{\left( 2L+1 \right) \left( 2\ell'+1 \right)}{4\pi\left( 2\ell+1 \right)} \right]^{1/2} \nonumber \\
                               & \times \left< L,0,\ell',2 | \ell,2 \right> \left< L,M,\ell',m'| \ell,m \right>.
\end{align}

In this paper we focus on mixing from the dominant mode, ($\ell' = 2, m' = 2$), with another $m=2$
mode, fixing these values. For these cases the Clebsch-Gordan selection rules require that $M=0$
and $\ell-2 \leq L \leq \ell+2$. Then our mixing coefficient takes the form
\beq
C_{\ell 2 2 2} = \sum_{L} b_{L0}\left[\frac{5 \left( 2L+1 \right)}{4\pi\left( 2\ell+1 \right)} \right]^{1/2} \left< L,0,2,2 | \ell,2 \right>^2.
\eeq

For example, complete expansions for $\ell=3$ and $\ell=4$ would yield
\begin{align*}
C_{3 2 2 2} &= \sqrt{\frac{5}{7\pi}} \frac{1}{132} \left( 22\sqrt{3} b_{10} + 33 \sqrt{5} b_{20} + 22 \sqrt{7} b_{30} \right. \\
            & \left. + 22 b_{40} + \sqrt{11} b_{50} \right),\\
C_{4 2 2 2} &= \sqrt{\frac{5}{\pi}} \frac{1}{4004} \left( 143\sqrt{5} b_{20} + 286\sqrt{7} b_{30} + 702 b_{40} \right. \\
            & \left. + 91\sqrt{11} b_{50} + 14 \sqrt{13} b_{60} \right).
\end{align*}

The shape of the distorted extraction sphere is determined by the coefficients $b_{L0}$: for real
$\alpha$, we need the $b_{L0}$ also to be real. The reality of the Clebsch-Gordan coefficients then
implies that  $C_{\ell 2 2 2}$ is also real.

\begin{figure}
\includegraphics*[width=3.5in]{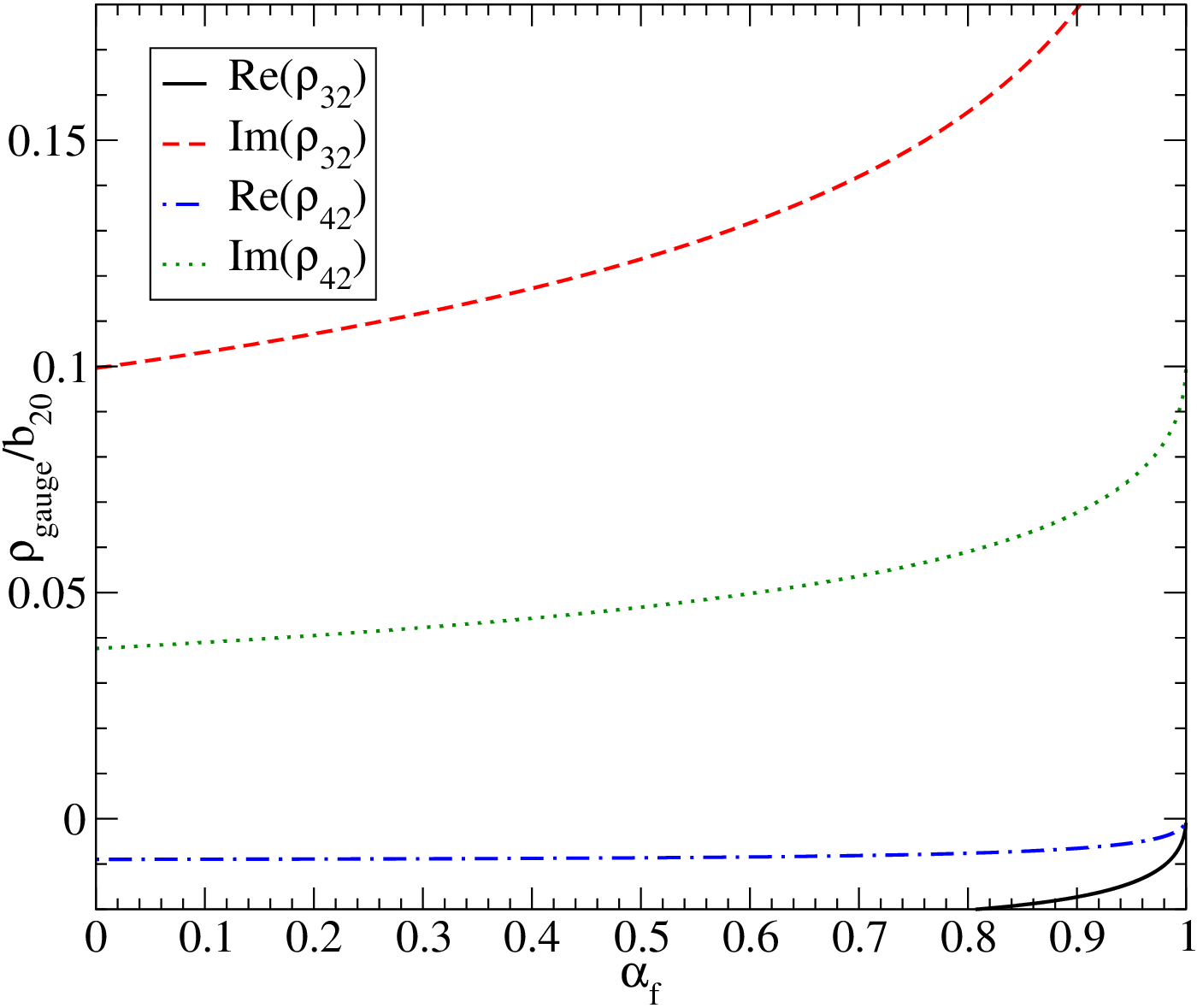}
\caption{Real and imaginary parts of $\rho_{{\rm gauge}, 32}$ and
$\rho_{{\rm gauge}, 42}$ (for $b_{20} = 1$) for post-merger Kerr BHs of
dimensionless spin $\alpf$ (and corresponding fundamental $(2,2)$ QNM frequency
$\Mf\sigtwotwoQNM$).}
\label{fig:leakage_ratios_BMS}
\end{figure}

The other ingredient in the waveform-mode perturbation \eqref{eq:mode_genperturb_BMS} is the
derivative with rrespect to $u$ on the right-hand side:
\[
\partial_u \psi_{\ell'm'}(u) = \partial_u \left( A(u) e^{i \varphi(u)}\right) = \left(\frac{\dot{A}}{A} + i \dot\varphi\right)\psi_{\ell'm'}(u).
\]
After merger, the effective coefficient $ \left(\dot{A}/A + i \dot\varphi\right)$ will asymptote to
a constant complex number:
\[
\left(\frac{\dot{A}}{A} + i \dot\varphi\right) \rightarrow - \frac{1}{\tau_{\ell'm'}} + i \omega_{\ell' m'} = i \sigma_{\ell' m'}.
\]
This implies a simple, QNM-driven leakage from the $(2,2)$ mode into higher-$\ell$ modes.
Collecting terms, and working with the strain-rate $\dot{h}$, during ringdown we have
\begin{align}
\dot{h}_{(\ell,2)}^{\rm gauge} &\approx \dot{h}_{(\ell,2)} + i C_{\ell 2 2 2} \, \sigma_{22} \, \dot{h}_{(2,2)} \nonumber \\
                               &  =     \dot{h}_{(\ell,2)} + \rho_{{\rm gauge}, \ell 2} \, \dot{h}_{(2,2)}.
\label{eq:mode_RDperturb_BMS}
\end{align}
In Fig.~\ref{fig:leakage_ratios_BMS}, we show the real and complex parts of the leakage parameters
$\rho_{{\rm gauge}, 32}$ and $\rho_{{\rm gauge}, 42}$ for the sweep of end-state spins $\alpf$,
assuming an unchanging scaling $b_{20} = 1$ (and all other $b_{L0} = 0$). The value of $b_{20}$ is
not physical, but gauge, and may differ between any two waveform determinations. The most important
property we note is that the BMS leakage coefficients are nearly pure-imaginary at any fixed
$b_{20}$ and any spin $\alpf$.

\subsection{Angular Basis Effects}
\label{ssec:basis_mixing}

Another possible path to mixing arises from considering what quasinormal-mode (QNM) frequencies
actually represent. QNMs were originally discovered in numerical black hole scattering studies
\cite{Vishveshwara:1970zz,Press:1971wr} and eventually understood as a key feature of the
perturbation theory of Kerr black holes \cite{Teukolsky:1972my}. In developing this theory,
Teukolsky worked with a background Kerr black hole in a very specific coordinate system due to
Boyer \& Lindquist \cite{Boyer:1966qh}.\footnote{Teukolsky theory can be reformulated on other
backgrounds; see, e.g., \cite{Campanelli:2000nc}.}

A perturbed Kerr black hole will ring down to quiescence through the emission of gravitational
waves. These waves will have characteristic frequencies $\omega_{\ell m}$ and damping times
$\tau_{\ell m}$ given by the hole's QNM spectrum.\footnote{We omit the principal quantum number
$n$, assuming that we are dealing with the slowest-damped fundamental ($n=0$) QNM.} While the
primary aim of QNM analysis is to determine the set of allowed complex frequencies
$\sigQNM \equiv \omega_{\ell m} + i/\tau_{\ell m}$, these frequencies are tied to the radial and
angular eigenfunctions arising from the separation of the perturbation equations. These angular
eigenfunctions are the spin-weighted \emph{spheroidal} harmonics,
$\SYmtwolm{\ell}{m}(\aw;\theta,\phi) \equiv \Smtwolm{\ell}{m}(\aw;\cos\theta) e^{i m \phi}$.
\footnote{Here we use the symbol $\sigma$ to denote a generic \emph{complex} frequency. $\sigQNM$
is a specific eigenvalue of the Kerr background.}

Numerical waveform extraction from binary mergers, on the other hand, typically decomposes the
waveforms onto the more generally motivated basis of spin-weighted \emph{spherical} harmonics
$\Ymtwolm{\ell}{m}(\theta,\phi)$, which correspond to a spheroidal harmonic basis with $\aw = 0$:
$\Ymtwolm{\ell}{m}(\theta,\phi) \equiv \SYmtwolm{\ell}{m}(0;\theta,\phi)$ \cite{Teukolsky:1972my}.
Buonanno \etal \cite{Buonanno:2006ui} demonstrated that using a spherical harmonic basis will
necessarily result in mixing of $(\ell, m)$ and $(\ell', m)$ quasinormal modes. Without an obvious
nontrivial choice for $\aw$ that  applies at all times, for all modes, over the course of the
evolving simulation, decomposing with $\aw \rightarrow 0$ seems a natural choice. Here we consider
an alternative choice, $\aw\rightarrow\Mf \alpf \sigtwotwoQNM$, hoping to limit much of the mode
mixing. Using this basis requires knowing the final Kerr state $(\Mf,\alpf)$ of the merger before
the decomposition can be applied, and the additional task of numerically computing the basis
functions (see Appendix~\ref{sec:spheroid_calc}). Still this basis is not optimal for the
subdominant modes. This unavoidable sub-optimality is discussed further in the next subsection.
The distinction between the spheroidal and spherical harmonics may be expected to yield the
appearance of mode-mixing in the numerical waveform results even if we have eliminated the gauge
freedom noted in the last section by optimal correspondence with a suitably perturbed
Boyer-Lindquist coordinate system.

To estimate the apparent mode-mixing from this basis mismatch, we can calculate the overlaps
between the spheroidal harmonics (for a particular $\aw$) and the spherical harmonics. That is, we
want to know the coefficients $s_{\ell' \ell m}$ in
\beq
\SYmtwolm{\ell}{m}(\aw;\theta,\phi) = \sum_{\ell'=2}^{\infty} s_{\ell' \ell m} \Ymtwolm{\ell'}{m}(\theta,\phi).
\label{eq:spheroidal_spherical_decomp}
\eeq

We describe our calculation of the $\SYmtwolm{\ell}{m}$ in Appendix~\ref{sec:spheroid_calc}.
To determine the overlaps $s_{\ell' \ell m}$, we decompose the properly normalized spheroidal
harmonic against the spherical harmonics in the usual way:
\begin{align*}
s_{\ell' \ell m} =& \oint d\Omega \SYmtwolm{\ell}{m}(\alpomtwotwo;\theta,\phi) \Ymtwolm{\ell'}{m}(\theta,\phi)^*\\
                 =& \int_{-1}^{1} dx \Smtwolm{\ell}{m}(\alpomtwotwo;x) \Smtwolm{\ell'}{m}(0;x)^*.
\end{align*}

\begin{figure}
\includegraphics*[width=3.5in]{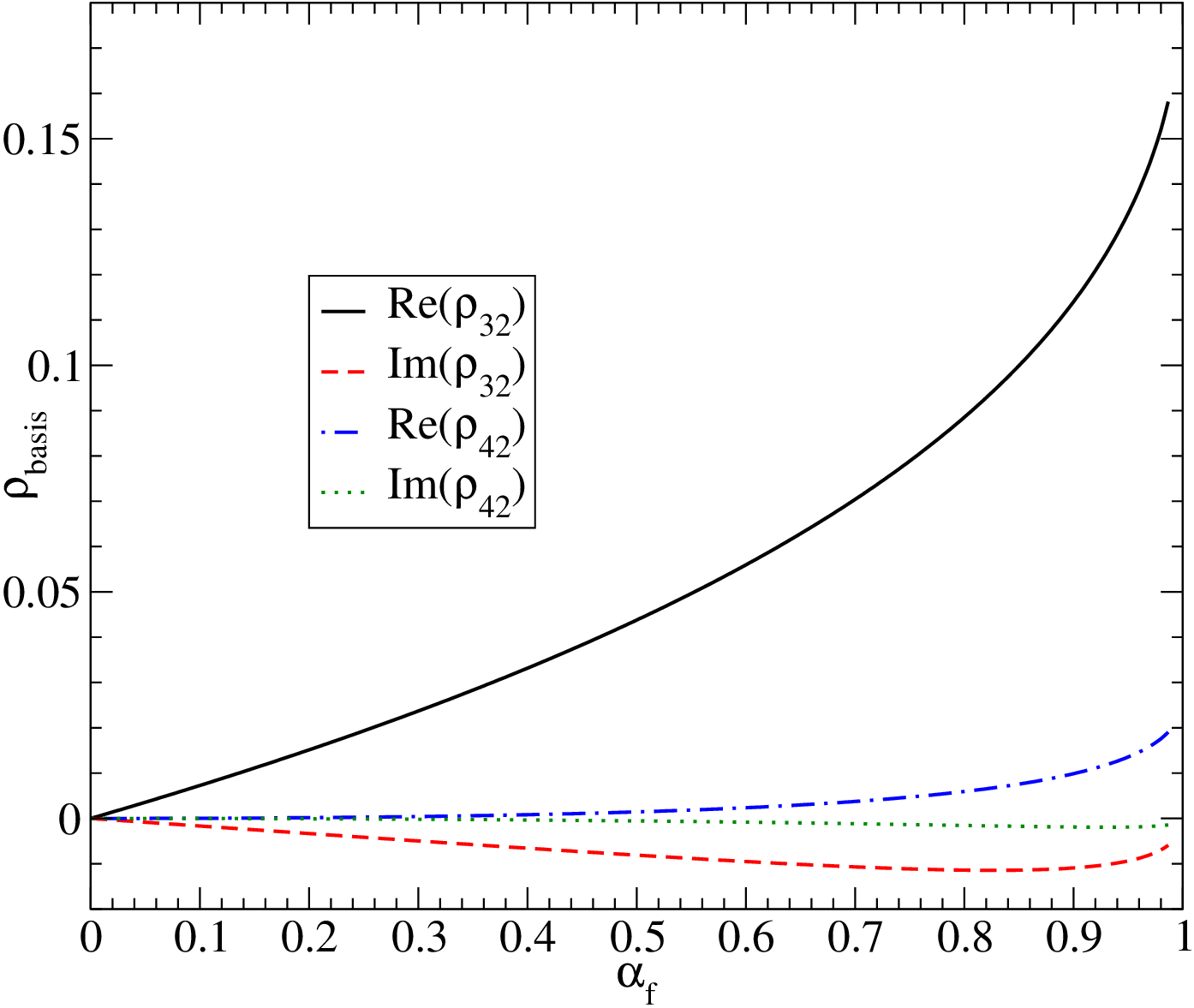}
\caption{Real and imaginary parts of the mixing coefficients
$\rho_{{\rm basis}, 32}$ and $\rho_{{\rm basis}, 42}$ for post-merger Kerr BHs
of dimensionless spin $\alpf$ (and corresponding fundamental $(2,2)$ QNM
frequency $\Mf\sigtwotwoQNM$). At the zero-spin limit $\alpf \rightarrow 0$,
the leakage vanishes.}
\label{fig:leakage_ratios}
\end{figure}

Now consider the idealized case where a physical ringdown signal is the simple
combination of the fundamental $(2,2)$, $(3,2)$, and $(4,2)$ quasinormal modes (we
omit $\aw$ arguments for brevity):
\begin{align}
\dot{h}(t,r,\theta,\phi) &= \sum_{\ell}^4 \mathcal{H}_{\ell 2}(t,r) \SYmtwolm{\ell}{2}(\theta,\phi) \nonumber \\
                         &\approx \sum_{\ell'}^4 \dot{h}_{\ell' 2}(t,r) \Ymtwolm{\ell'}{2}(\theta,\phi).
\end{align}
If we make the reasonable assumption that mixing $\ell \neq \ell'$ products can be ignored for
subdominant modes, then the measured spherical harmonic ringdown modes are approximately:
\begin{align}
\dot{h}_{(2',2)}^{\rm basis}(t,r)    &\approx s_{2' 2 2} \mathcal{H}_{22}(t,r), \nonumber\\
\dot{h}_{(\ell',2)}^{\rm basis}(t,r) &\approx s_{\ell' \ell 2} \mathcal{H}_{\ell 2}(t,r)  + \rho_{{\rm basis}, \ell 2} \dot{h}_{2'2}(t,r).
\end{align}
Here, the mixing coefficients are
\beq
\rho_{{\rm basis}, \ell 2} \equiv \frac{s_{\ell' 2 2}}{s_{2' 2 2}}.
\label{eq:spheroidal_mix_def}
\eeq
In Fig.~\ref{fig:leakage_ratios}, we plot the coefficients $\rho_{{\rm basis}, 32}$ and
$\rho_{{\rm basis}, 42}$, evaluated at $\aw = \alpomtwotwo$, where $\sigtwotwoQNM$ is the
fundamental QNM frequency of the $(2,2)$ mode for a Kerr hole of mass $\Mf$ and dimensionless spin
$\alpf$. Note that (a) there is no ambiguity in overall scale for these coefficients (unlike the
BMS-derived coefficients of the last section), and (b) they are strongly real-dominated.

We note here another manifestation of angular-basis mode-mixing demonstrated by Nu\~{n}ez \etal
\cite{Nunez:2010ra}. Those authors recast the Kerr perturbative problem using horizon-penetrating
coordinates and with a novel (non-Kinnersley) null tetrad. On this background, they were able to
show that the angular eigenfunctions are the (spin-weighted) \emph{spherical} harmonics. However,
the time-evolution of the radial mode functions for $(\ell,m)$ now involves the mode functions for
terms $(\ell \pm 1,m)$ and $(\ell \pm 2,m)$.

\subsection{Physical mixing}
\label{ssec:physical_mixing}

The discussion above exposes artifacts that arise from waveform decomposition using ordinary
spin-weighted spherical harmonic functions. Here we ask whether, even with extraction spheres in
the Boyer-Lindquist gauge, another decomposition using spin-weighted spheroidal harmonic functions
can avoid mode-mixing.  

The question is non-trivial. Although each leading-order quasinormal ringdown mode exhibits angular
dependence described by \emph{some} kind of spin-weighted spherical harmonic angular function, they
are not mutually given by the \emph{same} kind of spin-weighted spherical harmonic angular
functions, since each has its own distinct quasinormal frequency $\sigma_{\ell m}$, and
consequently a distinct preferred basis as labeled by $\aw = \Mf \alpf \sigma_{\ell m}$. We must
choose some particular orthonormal basis for the decomposition, and that basis cannot be
simultaneously optimal for each mode.

\begin{figure}
\includegraphics*[width=3.5in]{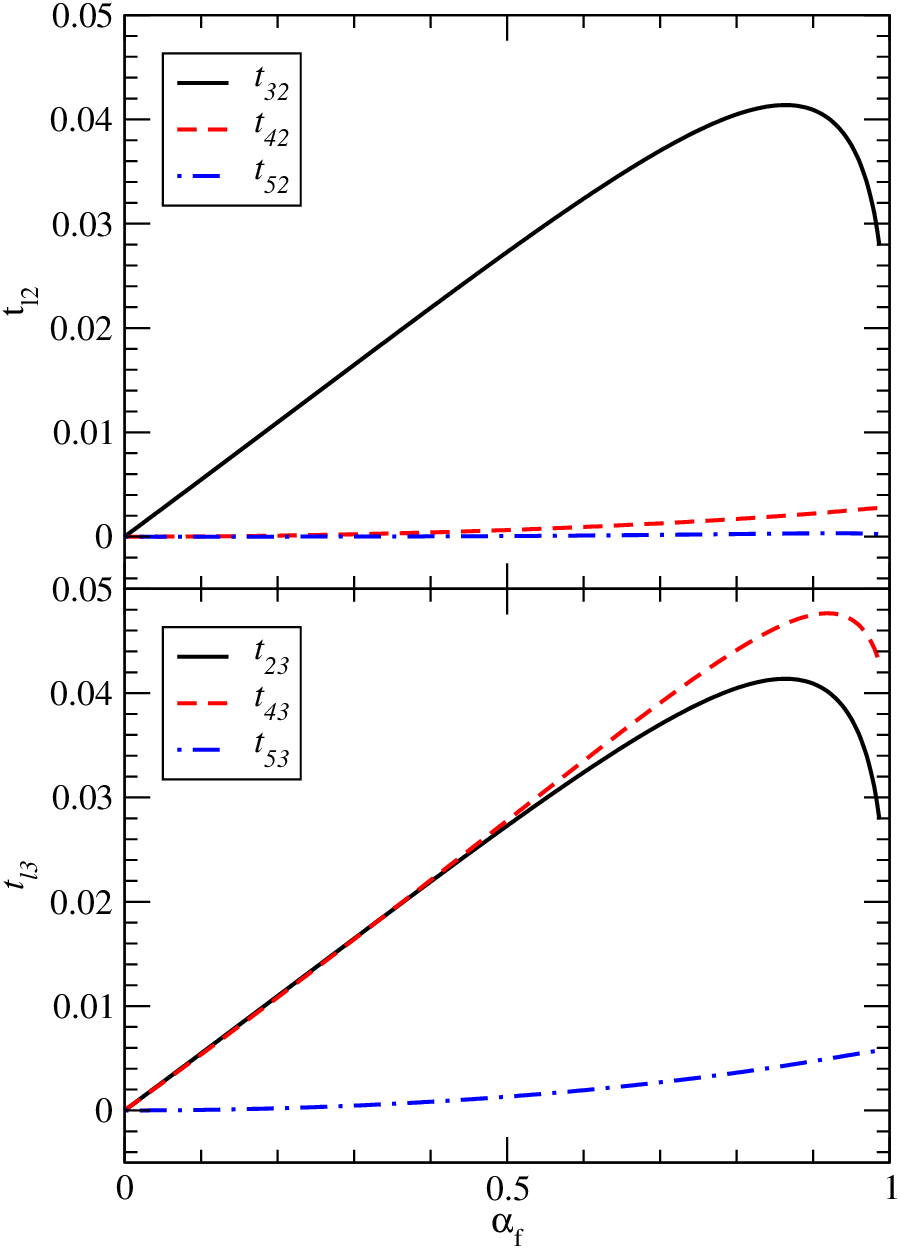}
\caption{[Top panel] Magnitude of overlap $t_{\ell 2}$ between
$\SYmtwolm{2}{2}$, evaluated at $\aw = \Mf \alpf \sigma_{22}$ and
$\SYmtwolm{\ell}{2}$, evaluated at $\aw = \Mf \alpf \sigma_{\ell 2}$.
[Bottom panel] Same for overlap $t_{\ell 3}$ between $\SYmtwolm{3}{2}$ and
$\SYmtwolm{\ell}{2}$. By definition, $|t_{32}| = |t_{23}|$.}
\label{fig:Slm_freq_mismatch}
\end{figure}

That the spheroidal harmonics associated with different QNM frequencies are not perfectly
orthogonal has been demonstrated for high-spin Kerr holes by Berti \etal \cite{Berti:2005gp}.
To quantify this for a general end-state spin $\alpf$, we define new overlaps $t_{\ell' \ell}$,
between spheroidal harmonics associated with different $m=2$ QNM frequencies:
\beq
t_{\ell' \ell} = \oint d\Omega \SYmtwolm{\ell}{2}(\Mf \alpf \sigma_{\ell 2}) \SYmtwolm{\ell}{2}(\Mf \alpf \sigma_{\ell' 2})^*.
\eeq
"The upper panel of Fig.~\ref{fig:Slm_freq_mismatch} shows the magnitude of these overlaps for 
$\ell = 2$ and several values of $\ell'$, while the lower panel shows the same for $\ell = 3$. 
From these plots, we see that the
spheroidal harmonics for different $\aw$ are not orthogonal, but show mixing by as much as
$\approx 4\%$ for high spins (though the maximum overlaps occur at sub-maximal spins, as noted by
\cite{Berti:2005gp}). The overlaps are also greatest for ``nearest neighbor'' modes:
$\ell = \ell' \pm 1$. For example, if we decomposed a waveform, including a non-trivial $(2,2)$
QNM, in the spheroidal basis corresponding to the $(3,2)$ mode ringdown frequency, then the
corresponding curve in Fig.~\ref{fig:Slm_freq_mismatch} would represent a mixing coefficient
analogous to those in the previous subsections. There is no choice of orthonormal basis that will
avoid all such mode mixing. In this sense, the angular non-orthogonality of the quasinormal mode
implies a form of physical mode-mixing, meaning that we can not perfectly isolate the QNM
frequencies by any choice of angular basis.

Fortunately it seems that the most evident mixing involves the dominant $(2,2)$ mode frequency
bleeding into higher-$\ell$ modes. With that assumption we may still eliminate most physical mixing
by choosing the basis compatible with this dominant quasinormal mode. If we decompose with the
basis labeled by $\aw=\alpomtwotwo$ then the orthogonality of this particular basis will completely
prevent the $(2,2)$ quasinormal mode from mixing into any other decomposed modal waveform
component. In this way we can eliminate any ``physical mixing'' of the particular form described in
Sec.~\ref{sec:bump_measured}.  Mixing among subdominant modes, or mixing of subdominant modes
into the decomposed $(2,2)$ waveform component will still occur at some level, but this is a
smaller effect, which we do not focus on in this paper.

%%%%%%%%%%%%%%%%%%%%%%%%%%%%%%%%%%%%%%%%%%%%%%%%%%%%%%%%%%%%%%%%%%%%%%%%%%%%%%%%%%%%%%%%%%%%%%%%%%%
%%%%%%%%%%%%%%%%%%%%%%%%%%%%%%%%%%%%%%%%%%%%%%%%%%%%%%%%%%%%%%%%%%%%%%%%%%%%%%%%%%%%%%%%%%%%%%%%%%%
%%%%%%%%%%%%%%%%%%%%%%%%%%%%%%%%%%%%%%%%%%%%%%%%%%%%%%%%%%%%%%%%%%%%%%%%%%%%%%%%%%%%%%%%%%%%%%%%%%%

\section{Simulations}
\label{sec:simulations}

\begin{figure*}
\includegraphics*[width=7.0in]{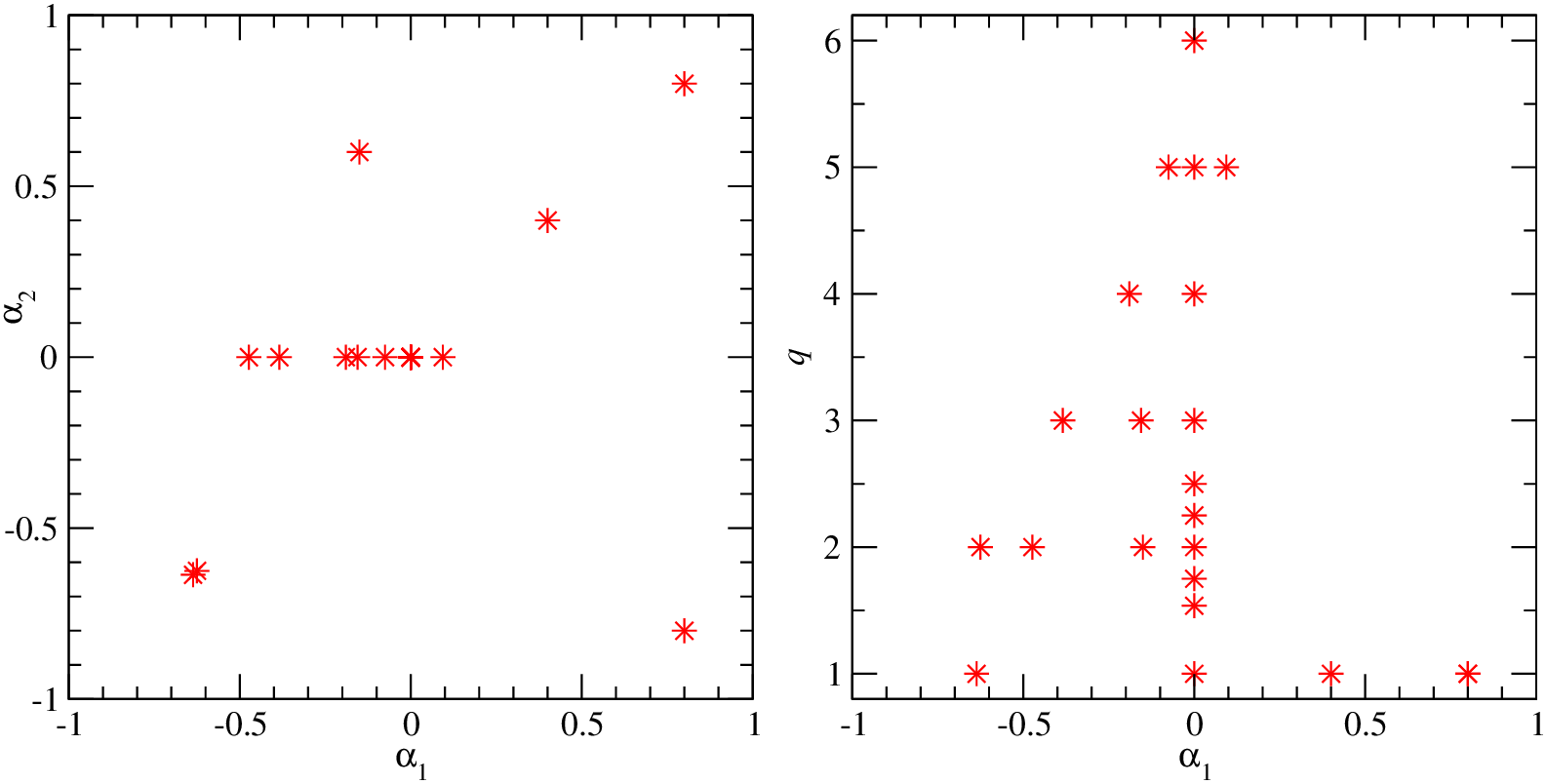}
\caption{Simulated configurations from Table~\ref{tab:equiv_ID}, represented as
points in two-dimensional $\{\alpha_1,\alpha_2\}$-space [left panel] and
$\{\alpha_1,q\}$-space [right panel]. Note that some points are associated with
multiple simulations.}
\label{fig:configuration_space}
\end{figure*}

To investigate the mixing in a systematic way, we have surveyed several existing simulations of
aligned-spin binaries, as well as carrying out new short simulations with the Goddard
\textsc{Hahndol} evolution code. We choose our new black-hole binary (BHB) configurations in
several groups of ``merger-equivalent'' classes, as described in Appendix~\ref{sec:equiv_ID}. The initial
parameters for all these simulations, old and new, are presented in Table~\ref{tab:equiv_ID}. In
Fig.~\ref{fig:configuration_space}, we show the distribution of these configurations as plots in
the two-dimensional configuration-spaces $\{\alpha_1,\alpha_2\}$ and $\{\alpha_1,q\}$, where
$q \equiv M_1/M_2>1$ is the mass ratio, and $\alpha_A \equiv S_A/M_A^2$ is the dimensionless spin
parameter of hole $A$, with physical values restricted to $\alpha_A \in [-1,1]$. Many of the longer
and higher-resolution evolutions have appeared in previous publications
\cite{Baker:2008mj,Kelly:2011bp}. Since our primary interest here is strictly in the late-merger
regime, newer evolutions begin only a few orbits before merger.

\begin{table*}\footnotesize
\caption{Physical and numerical parameters of the initial data for all the runs
presented. $m_{1,p}$ and $m_{2,p}$ are the bare puncture masses of the two
pre-merger holes. $r_0$ is the initial coordinate separation, while $P_{0t}$
and $P_{0r}$ are the initial transverse and radial components of the Bowen-York
linear momentum. $\MADM$ is the total energy of the initial data, while the
total infinite-separation mass of the system is estimated by the sum of the
initial Arnowitt-Deser-Misner (ADM) masses of the individual holes \cite{Baker:2002gm}. We have found
that for all cases here, this differs from the sum of apparent-horizon masses
(calculated at times between $t = 100$ and $200$), by less than a tenth of a
percent.}
\begin{tabular*}{0.75\textwidth}{c rrrrrrrrrrr}
\hline \hline
run name           & $m_{1,p}$    & $m_{2,p}$ & $S_{1z}$ & $S_{2z}$     & $r_0$        & $P_{0t}$        & $P_{0r}$        & $\MADM$  & $\sum_i M_{{\rm ADM}, i}$ \\
                   &              &           &          &              &              & $(\times 10^2)$ & $(\times 10^4)$ &          &                           \\
\hline
\texttt{X1\_UU}    & 0.301805     & 0.301805  &  0.2     &  0.2         &  8.20        & 10.32           &  0.00           & 0.988459 & 1.000804                  \\
\hline
\texttt{X1\_uu}    & 0.454575     & 0.454575  &  0.1     &  0.1         & 10.21        &  9.25           &  9.17           & 0.99223  & 1.002768                  \\
\hline
\texttt{X1\_00}    & 0.487231     & 0.487231  &  0.0     &  0.0         & 11.00        &  9.01           &  7.09           & 0.990514 & 1.000050                  \\
\texttt{X1\_UD}    & 0.301805     & 0.301805  &  0.2     & -0.2         & 11.00        &  9.01           &  7.09           & 0.990024 & 0.999222                  \\
\hline
\texttt{X1.5\_00}  & 0.581359     & 0.380645  &  0.0     &  0.0         &  7.12        & 11.75           & 29.17           & 0.987252 & 1.000000                  \\
\hline
\texttt{X1.75\_00} & 0.619237     & 0.345598  &  0.0     &  0.0         &  7.42        & 11.01           & 24.10           & 0.988129 & 1.000000                  \\
\hline
\texttt{X2\_00}    & 0.649344     & 0.314904  &  0.0     &  0.0         &  7.00        & 11.00           &  0.00           & 0.987939 & 1.000000                  \\
\texttt{X2\_DU}    & 0.648662     & 0.265507  & -0.066666667 &  0.066666667 & 10.00        &  8.52           &  7.63           & 0.990951 & 1.000009                  \\
\hline
\texttt{X2.5\_00}  & 0.699349     & 0.269501  &  0.0     &  0.0         &  7.40        &  9.79           & 20.53           & 0.989664 & 1.000000                  \\
\hline
\texttt{X3\_00}    & 0.738687     & 0.237505  &  0.0     &  0.0         &  8.88        &  7.88           &  8.96           & 0.991673 & 1.000000                  \\
\hline
\texttt{X4\_00}    & 0.790000     & 0.189000  &  0.0     &  0.0         &  8.47        &  6.96           &  0.00           & 0.992912 & 1.000310                  \\
\texttt{X5\_U0}    & 0.822007     & 0.157080  &  0.065083333 &  0.0     &  8.68        &  5.91           &  4.88           & 0.993733 & 1.000000                  \\
\texttt{X3\_d0}    & 0.731667     & 0.237705  & -0.087566063 &  0.0     &  9.06        &  7.84           &  8.76           & 0.99187  & 1.000000                  \\
\texttt{X2\_D0}    & 0.587677     & 0.317821  & -0.210380889 &  0.0     &  8.44        &  9.93           & 16.53           & 0.989967 & 1.000000                  \\
\texttt{X1\_DD}    & 0.390411     & 0.390411  & -0.159125  & -0.159125  & 11.98        &  8.84           &  1.20           & 0.990453 & 0.998786                  \\
\hline
\texttt{X5\_00}    & 0.824897     & 0.157031  &  0.0     &  0.0         &  8.67        &  5.97           &  5.85           & 0.993827 & 1.000000                  \\
\hline
\texttt{X6\_00}    & 0.848615     & 0.133064  &  0.0     &  0.0         &  7.55        &  5.84           &  6.94219        & 0.994008 & 1.000000                  \\
\texttt{X5\_D0}    & 0.822405     & 0.156318  & -0.052232639 &  0.0     &  8.09        &  6.32           &  7.00085        & 0.993556 & 1.000000                  \\
\texttt{X4\_D0}    & 0.778549     & 0.188766  & -0.1213184   &  0.0     &  8.57        &  7.04           &  7.78076        & 0.992926 & 1.000000                  \\
\texttt{X3\_D0}    & 0.692530     & 0.237756  & -0.21614625  &  0.0     &  9.17        &  7.93           &  8.80172        & 0.99219  & 1.000000                  \\
\texttt{X2\_DD}    & 0.531347     & 0.260245  & -0.277766667 & -0.069441667 & 10.72        &  8.56           &  7.81844        & 0.992008 & 1.000000                  \\
\hline
\hline \hline
\end{tabular*}
\label{tab:equiv_ID}
\end{table*}

\subsection{Numerics}
\label{ssec:numerics}

The initial momenta of the newer evolutions were chosen by integrating the post-Newtonian equations
of motion, as outlined in \cite{Husa:2007rh,Campanelli:2008nk}, with spin contributions to the
Hamiltonian adapted from
\cite{Buonanno:2005xu,Damour:2007nc,Porto:2006bt,Steinhoff:2007mb,Porto:2008jj,Steinhoff:2008ji},
and the flux from \cite{Blanchet:2006gy}. Note that we did not attempt to reduce the eccentricity
through tuning the initial momenta.

The new evolutions use the \textsc{Hahndol} code paired with the ``Curie''release of the Einstein
Toolkit \cite{Loffler:2011ay}, incorporating the Cactus Computational Toolkit \cite{cactus_web} and
the \textsc{Carpet} mesh-refinement driver \cite{carpet_web}.

In all cases, the initial data are of the standard Brandt-Br\"{u}gmann puncture type
\cite{Brandt:1997tf}, using the Bowen-York \cite{Bowen:1980yu} prescription for extrinsic curvature
that exactly satisfies the momentum constraint. We solve the remaining Hamiltonian constraint using
the \textsc{TwoPunctures} spectral code \cite{Ansorg:2004ds}.

To evolve these initial data, we employ the BSSNOK 3+1 decomposition of Einstein's vacuum equations
\cite{Nakamura:1987zz,Shibata:1995we,Baumgarte:1998te}, with the alternative conformal variable
$W \equiv e^{-2 \phi}$ suggested in \cite{vanMeter:2006g2n,Tichy:2007hk,Marronetti:2007wz},
constraint-damping terms suggested in \cite{Duez:2004uh}, and the dissipation terms suggested in
\cite{Kreiss73,Hubner:1999th}. Our gauge conditions are the specific 1+log lapse and Gamma-driver
shift described in \cite{vanMeter:2006vi}, which constitute a variant of the now-standard ``moving
punctures'' approach \cite{Campanelli:2005dd,Baker:2005vv}. Our spatial derivatives use
sixth-order-accurate differencing stencils, with the exception of advection derivatives, where we
use fifth-order-accurate mesh-adapted differencing (MAD) \cite{Baker:2005xe}. Our time-integration
is performed with a fourth-order Runge-Kutta algorithm.

\subsection{Waveform Extraction}
\label{ssec:extraction}

We extract the gravitational waveforms from the simulations through the radiative Weyl scalar
$\psi_4$ \cite{Baker:2001sf}. This is evaluated throughout the grid, and interpolated onto a set of
coordinate spheres at extraction radii $\Rext \in [40M, 90M]$. Over each sphere, the interpolant is
integrated against the set of spin-weighted spherical harmonics $\Ymtwolm{\ell}{m}(\theta,\phi)$,
up to $\ell=5$.

In the extraction region, the grid spacing is between $M/2$ and $2M$, depending on the central
resolution of the simulation. This is generally too coarse to resolve higher-frequency (and
higher-$m$) modes with accuracy. Even for the dominant, relatively low-frequency, $(2,\pm 2)$
modes, dissipation effects are visible that spoil the $1/\Rext$ extrapolation near and after
merger. For this reason, we have used an $\Rext$-extrapolation scheme that includes an explicit
dissipative term in the amplitude of each mode:
\[
A_{\ell m}(\Rext) = a_{-1} \Rext + a_0 + a_2 \Rext^{-2} , \;\; \varphi_{\ell m}(\Rext) = b_0 + b_2 \Rext^{-2}.
\]
We have found this extrapolation procedure to be robust only for the higher-resolution simulations
in this paper.

As a result, a waveform-derived quantity $f$ will have errors due to finite extraction radius and
finite resolution. For this paper, we make a very conservative error estimate by adding
uncertainties linearly:
\[
\Delta f = \Delta_{\Rext} f + \Delta_{h} f.
\]
For the finite-$\Rext$ error, we assume an uncertainty equal to the difference between the
coefficient from the $\Rext$-extrapolated highest-resolution data and that measured from the
largest finite-$\Rext$ data at the same resolution. For finite-resolution error, we use the
difference between the same-extraction-radius data at the coarse and fine resolutions as our
estimate of the error in the fine-resolution result. For many configurations, we only have a single
resolution available and the \Rext-extrapolation is not reliable at this resolution. For these, we
adopt a conservative overall error estimate by taking the average error from comparable
two-resolution configurations\footnote{By ``comparable'', we mean configurations that used the same
numerical executable and grid structure, and whose lower-resolution version matched that of the
single-resolution configurations.} and multiply it by 1.5. For amplitudes, this is a relative
error, while for phase measurements, it is the absolute error.

%%%%%%%%%%%%%%%%%%%%%%%%%%%%%%%%%%%%%%%%%%%%%%%%%%%%%%%%%%%%%%%%%%%%%%%%%%%%%%%%%%%%%%%%%%%%%%%%%%%
%%%%%%%%%%%%%%%%%%%%%%%%%%%%%%%%%%%%%%%%%%%%%%%%%%%%%%%%%%%%%%%%%%%%%%%%%%%%%%%%%%%%%%%%%%%%%%%%%%%
%%%%%%%%%%%%%%%%%%%%%%%%%%%%%%%%%%%%%%%%%%%%%%%%%%%%%%%%%%%%%%%%%%%%%%%%%%%%%%%%%%%%%%%%%%%%%%%%%%%

\section{Analysis of Waveforms}
\label{sec:analysis}

Using the ringdown data from all the simulations in Sec.~\ref{sec:simulations}, we performed
least-squares fits to the real part of the strain-rate $(2,2)$ and $(3,2)$ waveforms, using the
forms of Eqs.~\eqref{eq:22mod_QNM_model}-\eqref{eq:32mod_QNM_model}. Our fit is over the
window $t \in [20,55]$, where $t=0$ is the time of peak $(2,2)$ mode amplitude. By starting $20M$
after peak amplitude, we ensure that we are in the linear ringdown regime; by stopping at $55M$, we
avoid the low-amplitude degradation seen in late-ringdown waveforms. As the tabulated version of
the results would be excessively long, we present our raw results purely graphically.

We begin by showing the nature of the complex numerical ``leakage parameter'' derived from the
ratio of fitted parameters from the measured $(2,2)$ and $(3,2)$ modes during ringdown, using
\eqref{eq:22mod_QNM_model}-\eqref{eq:32mod_QNM_model}:
\beq
\rho_{{\rm num}, 32} \equiv \frac{\rho_{32} A_{22} e^{i(\sigma_{22} t + \delta_{22})}}{A_{22} e^{i(\sigma_{22} t + \delta_{22})}}.
\label{eq:rhonum_def}
\eeq
Figure~\ref{fig:QNM_fit_32_ratio} shows the real and imaginary parts of this leakage for all
configurations presented in this paper, as a function of the dimensionless spin $\alpf$ of the
post-merger hole. 

\subsection{Comparing Hypotheses}
\label{ssec:hypoth_test}

\begin{figure}
\includegraphics*[width=3.5in]{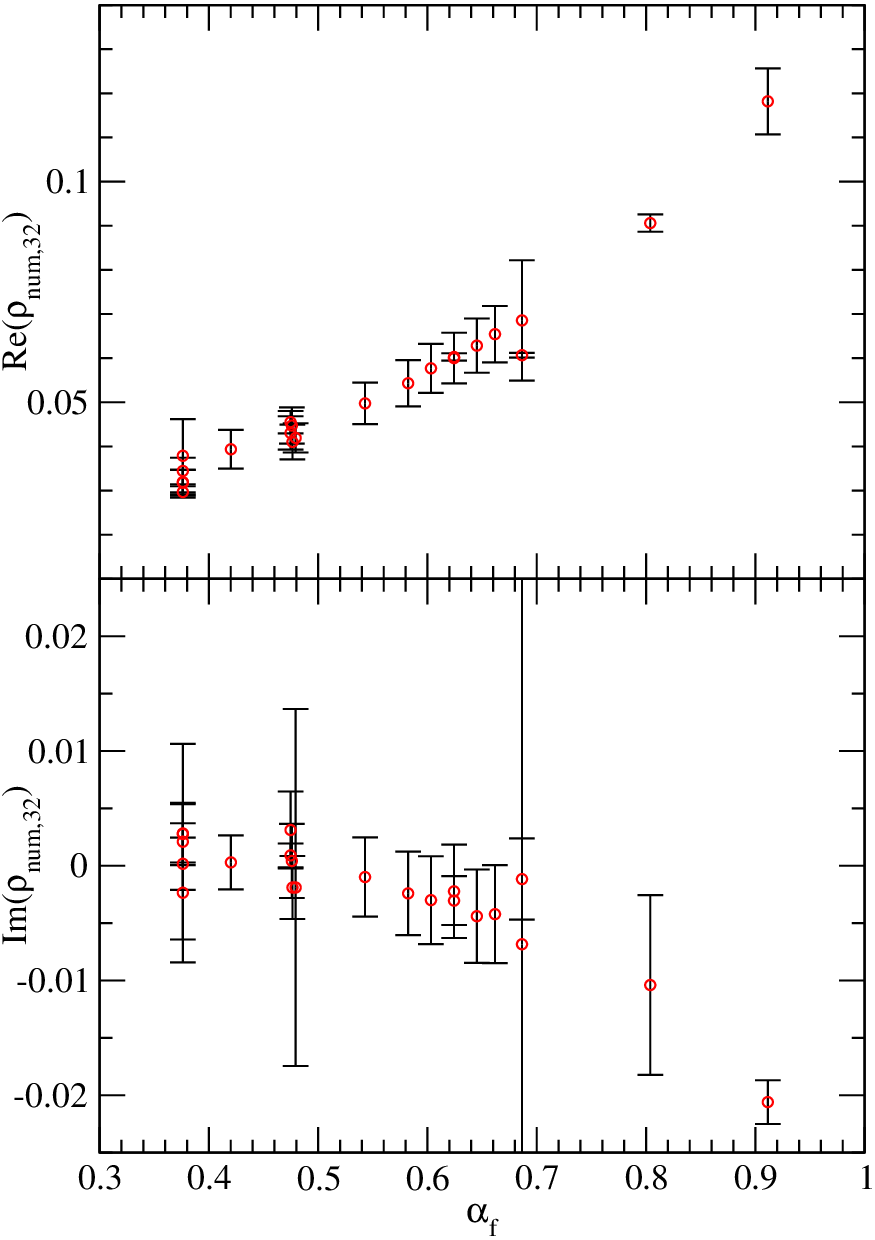}
\caption{Real [top panel] and imaginary [bottom panel] parts of complex leakage
parameter $\rho_{{\rm num}, 32}$ \eqref{eq:rhonum_def}, shown as a function of
$\alpf$, the dimensionless spin of the post-merger hole for all runs,
indicating that in almost all cases, the leakage parameter is predominantly
real.}
\label{fig:QNM_fit_32_ratio}
\end{figure}

In Sec.~\ref{sec:mode_mix_model} we discussed two possible causes for mode-mixing
effects of the form
\beq
\dot{h}_{(\ell,2)}^{\rm model} = A_{\ell2} e^{i(\sigma_{\ell2} t + \delta_{\ell2})} + \rho_{\ell2} A_{22} e^{i(\sigma_{22} t + \delta_{22})}, \label{eq:l2mod_QNM_model}
\eeq
described in Sec.~\ref{sec:bump_measured}. If the mixing is caused by BMS supertranslation gauge
ambiguity, then we would expect nearly pure imaginary $\rho_{\ell2}$.  On the other hand, if the
mixing derives from the distinction between spheroidal and spherical harmonic angular functions,
then we expect predominantly real $\rho_{\ell2}$ of a quantified size. In
Fig.~\ref{fig:QNM_fit_32_ratio} we see that the argument of $\rho_{{\rm num},3 2}$ is close to
zero, within error bars for most cases, making $\rho_{{\rm num}, 32}$ predominantly real,
consistent with the spheroidal harmonic hypothesis. The largest deviations from zero are also those
with the largest uncertainties arising from the QNM fit process.

The analysis of Sec.~\ref{ssec:gauge_mixing} suggests that a change of supertranslation gauge
would give rise to mode-mixing coefficients with a numerically significant imaginary part in the
measured waveform. Since the imaginary part of $\rho_{{\rm num},3 2}$ is so small, any
supertranslation gauge effects are negligible at the level of interest here. We can estimate the
degree of gauge constraint implied by a null measurement of this effect. From the bottom panel of
Fig.~\ref{fig:QNM_fit_32_ratio}, one sees that the imaginary component of the mixing coefficient
$\rho_{{\rm num},3 2}$ is constrained to values within $\pm 0.02$ in almost all cases. If we
generously assumed that all of this imaginary mixing was caused by gauge distortion of the
extraction sphere, by comparison with Fig.~\ref{fig:leakage_ratios_BMS}, we would conclude that the
amplitude of the distortion ($b_{20}$ specifically) would have to be smaller than about $0.2M$,
suggesting a remarkable level of supertranslation gauge optimality in these simulations. 

\subsection{Testing the Spheroidal Leakage Model}
\label{ssec:spheroid_test}

\begin{figure}
\includegraphics*[width=3.5in]{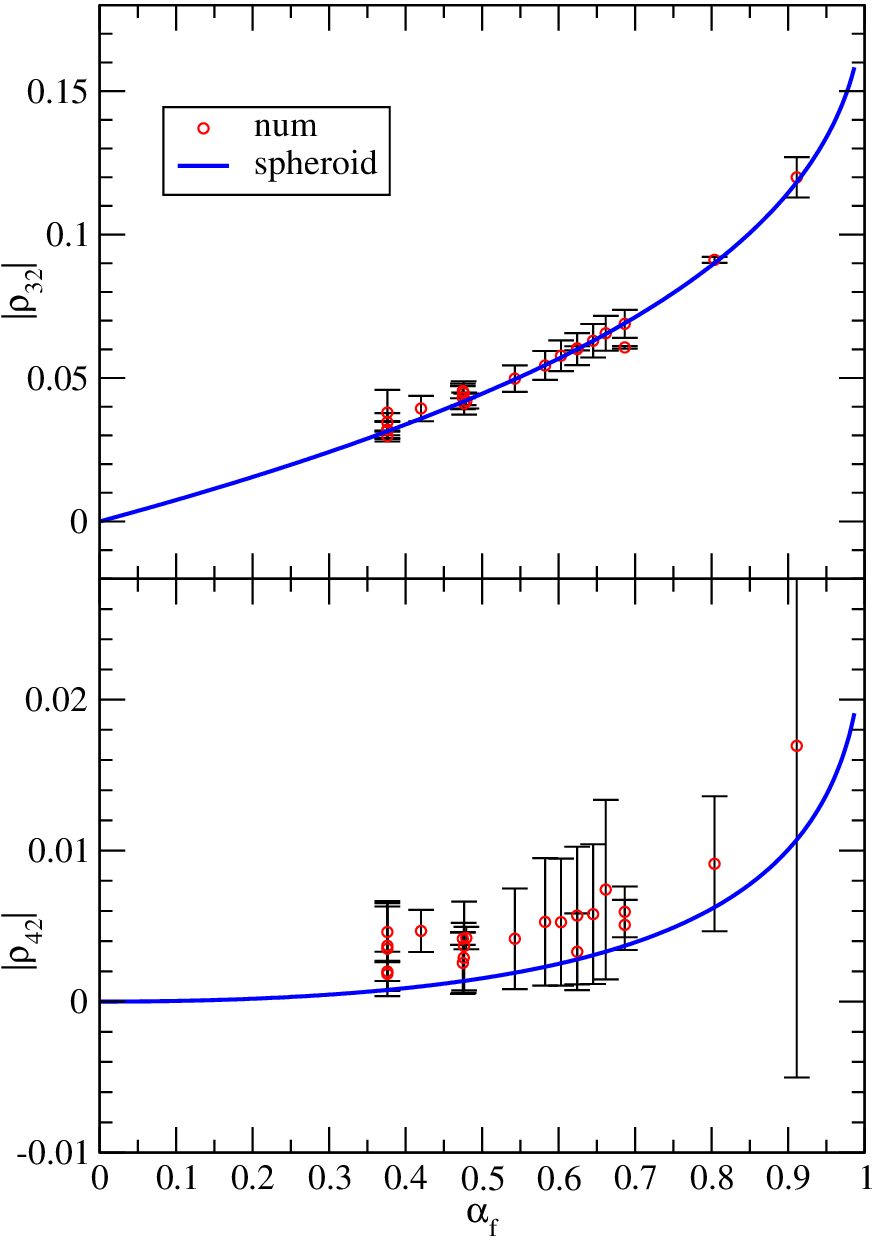}
\caption{[Top panel] Amplitude of measured $(2,2)\rightarrow(3,2)$ leakage
parameter $\rho_{{\rm num}, 32}$ as a function of $\alpf$, the dimensionless
spin of the post-merger hole for all runs. The curve is the theoretical ratio
$\rho_{{\rm basis}, 32}$ due to spheroidal-spherical leakage. [Bottom panel]
Same for $(2,2)\rightarrow(4,2)$ leakage ratio $\rho_{{\rm num}, 42}$.}
\label{fig:amp22_peak_ratio}
\end{figure}

We have seen that the numerical results for the complex argument of $\rho_{32}$ are consistent with
the spherical-spheroidal mixing hypothesis, but this hypothesis also makes quantitative predictions
for the magnitude $\left|\rho_{32}\right|$. In the top panel of Fig.~\ref{fig:amp22_peak_ratio}, we
plot the magnitude of $\rho_{{\rm num}, 32}$ as a function of $\alpf$. We overlay these points with
the magnitude of the leakage coefficients $\rho_{{\rm basis}, 32}$ \eqref{eq:spheroidal_mix_def}
plotted in Fig.~\ref{fig:leakage_ratios} (blue solid curve). From the close fit, it appears that
the leakage is in fact dominated by this spheroidal/spherical harmonic mismatch. That is, even
though the post-merger background coordinate system should not be expected to closely resemble the
Kerr-Boyer-Lindquist slicing assumed by Teukolsky's perturbative work, nevertheless, this expected
warping is not as important as our choice of harmonic basis functions.

The bottom panel of Fig.~\ref{fig:amp22_peak_ratio} shows the complex amplitude of the equivalent
parameter $\rho_{{\rm num}, 42}$ governing the leakage of the $(2,2)$ mode into the measured
$(4,2)$ mode. Although this is also consistent with expectations from angular-basis mixing (blue
solid curve), the relative errors swamp the numerical data, and higher-resolution numerics will be
needed to establish the relation unambiguously.

\subsection{Finding the Residual $(3,2)$ Mode Amplitude}
\label{ssec:residual}

If we regard the measured $(3,2)$ mode as the combination of a ``true'' $(3,2)$ mode
$A_{3 2} \exp {i(\sigma_{3 2} t + \delta_{\ell2})}$ and a piece of the $(2,2)$ mode, we may ask
whether we can model the residual $(3,2)$ contribution. When looking at the entire suite of
simulations, it is difficult to see a distinct pattern in these true $(3,2)$ amplitudes. However,
it is instructive to carry out a particular slice in configuration space.

\begin{figure}
\includegraphics*[width=3.5in]{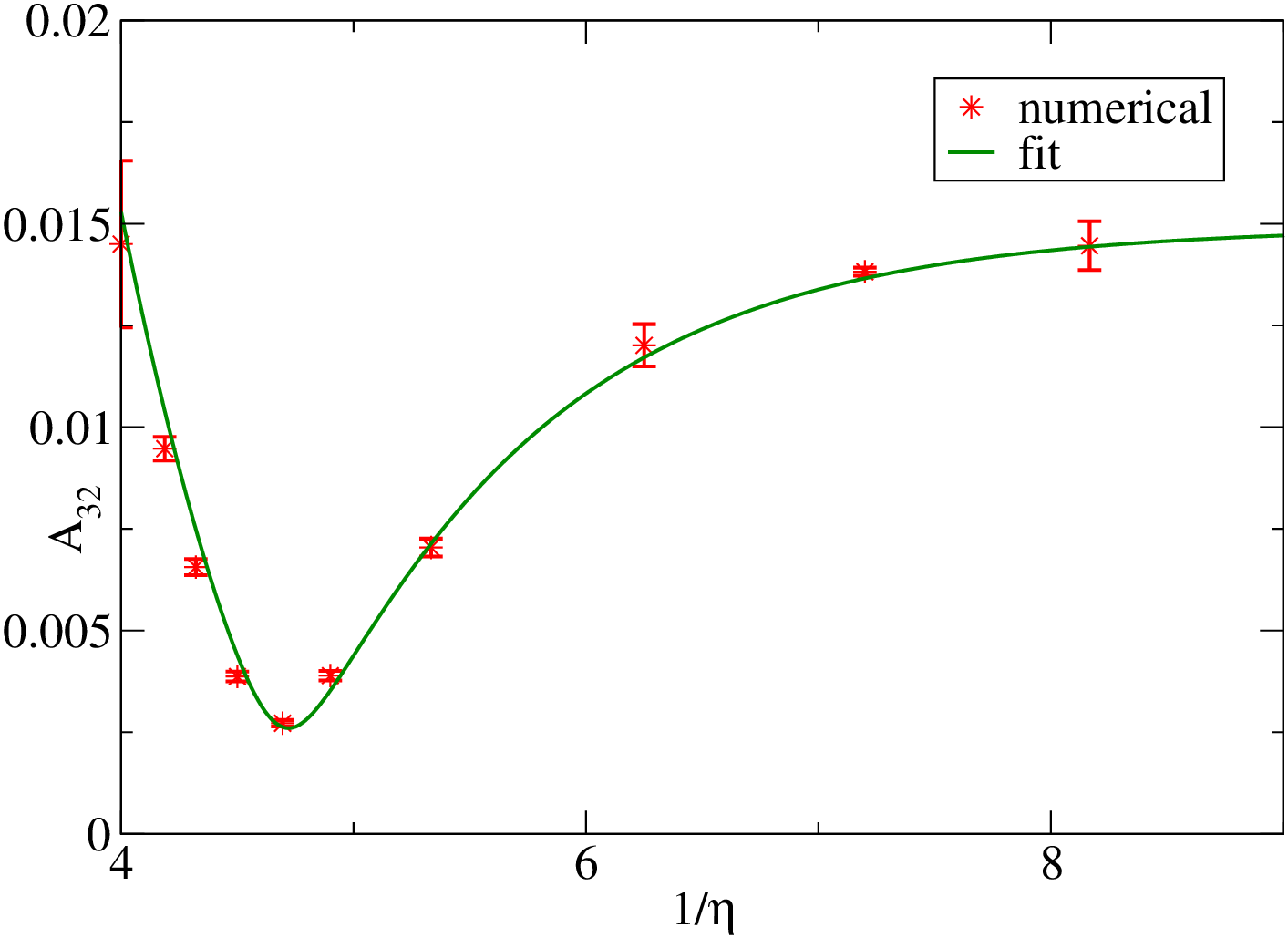}
\caption{Corrected amplitude $A_{32}$ of the $(3,2)$ strain-rate mode for
nonspinning binaries with mass ratio $q \in \{ 1.0, 6.0 \}$. The solid (green)
curve is a fit to these points of the functional form \eqref{eq:A32_fit}.}
\label{fig:amp32_nonspinning}
\end{figure}

In Fig.~\ref{fig:amp32_nonspinning}, we show a subset of the $(3,2)$ amplitudes formed by the
mergers of nonspinning binaries, with mass ratio $q \equiv M_1/M_2 \in \{ 1.0, 6.0 \}$. Error bars
in this plot have been estimated in the same way as for Fig.~\ref{fig:amp22_peak_ratio}. Clearly
the high-$q$ behavior seems to decay to some constant amplitude, while there is some local minimum
around $\eta = 0.21$ (between $q=2$ and $q=2.25$), indicating that perhaps at this mass-ratio, the
$(3,2)$ QNM is hardly excited at all.

We also present an empirical fit to this data of the functional form
\beq
A_{32}(\eta) = \sqrt{\left(a - b e^{-\lambda/\eta}\right)^2 + c^2 }, \label{eq:A32_fit}
\eeq
where the parameters take the values $a = 0.0147 \pm 0.0002$, $b = 1.5 \pm 0.1$,
$c = 0.0026 \pm 0.0002$, and $\lambda = 0.98 \pm 0.02$. Fits of this form are expected to be useful
for generating merger template waveforms for the subdominant modes.

%%%%%%%%%%%%%%%%%%%%%%%%%%%%%%%%%%%%%%%%%%%%%%%%%%%%%%%%%%%%%%%%%%%%%%%%%%%%%%%%%%%%%%%%%%%%%%%%%%%
%%%%%%%%%%%%%%%%%%%%%%%%%%%%%%%%%%%%%%%%%%%%%%%%%%%%%%%%%%%%%%%%%%%%%%%%%%%%%%%%%%%%%%%%%%%%%%%%%%%
%%%%%%%%%%%%%%%%%%%%%%%%%%%%%%%%%%%%%%%%%%%%%%%%%%%%%%%%%%%%%%%%%%%%%%%%%%%%%%%%%%%%%%%%%%%%%%%%%%%

\section{Discussion}
\label{sec:discuss}

In this paper, we have investigated ``bumps'' measured in the merger-ringdown portion of certain
gravitational-radiation angular waveform modes from the numerical simulation of the coalescence of
black-hole binaries (BHBs). These bumpy modes appear to contain significant contributions from the
dominant $(2,2)$ mode, indicating some kind of mode-mixing at work.

We have considered three classes of effects that may contribute to mode-mixing in numerically
extracted and decomposed merger-ringdown waveforms. These are: gauge effects, arising from
supertranslation gauge freedom for outgoing radiation in general asymptotically flat spacetimes
(see Sec.~\ref{ssec:gauge_mixing}); angular-basis effects, relating to a choice between
spin-weighted spherical or quasinormal-mode-adapted spheroidal harmonic bases
(Sec.~\ref{ssec:basis_mixing}); and physical quasinormal-mode mixing effects that are independent
of any representation changes (Sec.~\ref{ssec:physical_mixing}).

We have identified and analyzed the measured mode-mixing bumps in the most prominent of the bumpy
gravitational waveform modes modes --- $\ell = 3$,$m = 2$ --- measured from a set of numerical
evolutions of aligned-spin BHB mergers. Our analysis has allowed us to distinguish between the
contributions of our three mode-mixing effects.  We find that the angular-basis effects dominate.
Although other kinds of effects may be present -- like the frequency-dependent gauge
supertranslations discussed in Sec.~\ref{ssec:gauge_mixing} -- they cannot be seen clearly here
with the level of accuracy available from our current simulations.

In this way our analysis further codifies the results from the ringdown stage of the aligned-spin
mergers. This was originally prompted by our work on a multi-mode waveform model based on the
\emph{implicit rotating source} (IRS) picture of black-hole merger
\cite{Baker:2008mj,Kelly:2011bp}. In this model, the dominant and leading subdominant waveform
modes from binary mergers were seen to share a common \emph{rotational phase}, with a corresponding
rotational frequency that increased monotonically through inspiral and merger, reaching a plateau during
ringdown. The corresponding mode amplitudes could be modeled by a simple, few-parameter functional
form that depends on the frequency function, with a single well-defined peak. Attempting to extend
this to the $(3,2)$ mode proved problematic, as the measured mode was no longer monotonic in
frequency, or single-peaked in amplitude.

More broadly, we expect our results to provide guidance in the ongoing effort of combining results
of analytic and numerical relativity studies toward the goal of a fully developed family of
efficient and accurate black-hole merger waveforms. Because the comparison of waveform models is
typically conducted mode by mode in decomposed form, the issues we have studied may lead to
unnecessarily spurious features in particular waveform representations.

We estimate, for instance, that supertranslation gauge changes that would effectively distort the
shape of arbitrarily large waveform-extraction spheres on scales of order $\Mf$ or smaller would be
sufficient to qualitatively influence the mode-mixing features focused on in this study. The
absence of such effects is itself intriguing, suggesting that we have achieved nearly optimal
choice of supertranslation gauge. Our near-optimal spheroidal harmonic basis is consistent with
quasinormal-mode distortions of Kerr space-time in the Boyer-Lindquist coordinate system. That we
see negligible supertranslation mode-mixing suggests that the outer regions of our numerical
space-times asymptotically  approach distorted Kerr in Boyer-Lindquist coordinates faster (in
powers of $1/r$) than the asymptotic approach to perturbed Minkowski spacetime. This seems
plausible, based on our choice of numerical gauge, which approximates maximal time slicing and
$\tilde\Gamma^i=0$ spatial coordinates. The latter condition will yield spatially isotropic
coordinates where possible.

Nonetheless, it seems that we have been lucky to stumble onto a near-optimal representation as
other incompatible gauge choices may also be reasonable in the numerical simulation context. In
continued pursuit of higher-precision waveform comparisons and higher-fidelity analytic models
(see, e.g., the NR-AR project \cite{NRARwebsite}), we expect such considerations to grow in
significance. (They may also be crucial in studies of how the pre-merger BHB configuration is
encoded in the relative amplitude of different quasinormal modes during ringdown; see, e.g.
\cite{Kamaretsos:2012bs}.) Similarly we find that physical mode-mixing among the quasinormal modes
will prevent any orthonormal representation from fully separating frequencies at sufficiently high
precision.

For simulations similar to ours, where gauge and physical mixing effects remain small, and the
primary source of mixing involves the $(2,2)$ mode, our results suggest that decomposition with a
spheroidal harmonic basis $\{\SYmtwolm{\ell}{m}(M_f\alpha_f\sigma_{22};\theta,\phi)\}$ may be close
to an optimal basis for achieving modal frequency separation, and thus nearly beat-free waveforms.

It may be asked whether the conclusions drawn here can be applied to the pre-merger waveform
signal. We know that the PN mode amplitudes (see, for instance, Eqs. (4.17) of \cite{Arun:2008kb})
are dominated by the $(2,\pm 2)$ (quadrupole) spherical harmonic modes, with $(\ell > 2,\pm 2)$
modes entering at higher PN order. It might be possible, in principle, to find a ``best possible''
effective background spin parameter $\alpha_{\rm eff}$ whose associated spheroidal harmonic basis
would absorb most of these higher-$\ell$ modes; in practice, however, this would be numerically
impractical at any fixed frequency, and of course, the frequency would change continuously during
inspiral, as (presumably) would the spin, since the binary is constantly losing angular momentum.

\acknowledgments

The new numerical evolutions performed for this paper were carried out on the machine
\emph{Pleiades} at NASA's Ames Research Center. The work was supported by NASA grant
09-ATP09-0136. The authors would like to thank Enrico Barausse, Emanuele Berti, Alessandra
Buonanno, Rafael Porto, Luciano Rezzolla, Jeremy Schnittman, and James van Meter for useful
comments.

\appendix

\section{Calculating Spheroidal Harmonics}
\label{sec:spheroid_calc}

As there are no closed-form solutions for the $\SYmtwolm{\ell}{m}$, we must proceed numerically.
While setting up the popular continued-fraction method for computing the quasinormal-mode (QNM)
frequencies of a Kerr black hole, Leaver \cite{Leaver:1985ax} presents the following power-series
expansion for the polar-angle function $\Smtwolm{\ell}{m}(\aw;\cos\theta)$, due originally to Baber
\& Hass\'e \cite{BaberHasse_1935} (we specialize here to $s=-2$):
\begin{align}
\Smtwolm{\ell}{m}(\aw;x) =& e^{\aw x} (1+x)^{|m+2|/2} (1-x)^{|m-2|/2} \nonumber \\
                          & \times \sum_{n=0}^{\infty} a_n (1+x)^n, \label{eq:spheroidal_series_sol}
\end{align}
where the expansion coefficients $a_n$ are determined up to an overall scaling --- the value of
$a_0$ --- by the same recurrence relations that yield the QNM frequencies. For our desired Kerr
spin $\alpf$, we first determine the (complex) fundamental QNM frequency of the $(2,2)$ mode,
$\Mf\sigtwotwoQNM$. Next, assuming $a_0=1$, we use the recurrence relations from
\cite{Leaver:1985ax} to determine the $a_n$ (in practice, we truncate the series at $n=14$).
Requiring that
\[
\int_{-1}^{1} dx \left| \Smtwolm{\ell}{m}(\alpomtwotwo;x) \right|^2 = 1
\]
then fixes $a_0$, supplying the correct normalization of the $a_n$.

%%%%%%%%%%%%%%%%%%%%%%%%%%%%%%%%%%%%%%%%%%%%%%%%%%%%%%%%%%%%%%%%%%%%%%%%%%%%%%%%%%%%%%%%%%%%%%%%%%%
%%%%%%%%%%%%%%%%%%%%%%%%%%%%%%%%%%%%%%%%%%%%%%%%%%%%%%%%%%%%%%%%%%%%%%%%%%%%%%%%%%%%%%%%%%%%%%%%%%%
%%%%%%%%%%%%%%%%%%%%%%%%%%%%%%%%%%%%%%%%%%%%%%%%%%%%%%%%%%%%%%%%%%%%%%%%%%%%%%%%%%%%%%%%%%%%%%%%%%%

\section{Kerr-Equivalent Black-Hole Binaries}
\label{sec:equiv_ID}

The end-point of any merger of BHBs in vacuum is expected to be a single Kerr black hole,
parametrized by two numbers, the mass $\Mf$ and spin angular momentum
$\vec{S}_{\rm f} = \alpf \Mf^2$. These should satisfy the global conservation rules:
\begin{align}
\Mf             &= \MADM - E_{\rm rad}, \\
\vec{S}_{\rm f} &= \JADM - \vec{J}_{\rm rad},
\end{align}
where $\MADM$ and $\JADM$ are the Arnowitt-Deser-Misner (ADM) energy and total angular momentum of
the initial data, and $E_{\rm rad}$ and $\vec{J}_{\rm rad}$ are the energy and angular momentum
emitted in gravitational radiation during the course of the evolution.

Fixing the initial separation of the binary, and taking its total mass to be $M = M_1 + M_2$
($> \MADM$ for any finite initial separation), and assuming zero eccentricity, the black-hole binary
will have seven free parameters: $\{q, \vec{S}_1, \vec{S}_2\}$, where $q \equiv M_1/M_2 > 1$ is the
mass ratio, and $\vec{S}_A$ are the spin angular momentum vectors of the two holes. However, the
end-state has just two parameters: $\{ \Mf, \vec{S}_{\rm f} \}$, so there must be a large degeneracy
in the initial parameters.

Viewing the BHB coalescence as a kind of simple particle interaction, Boyle \etal \cite{Boyle:2007ru}
used symmetry arguments to restrict the possible end-states of the BHB merger. This is the basis of
end-state formulae by Tichy \& Marronetti~\cite{Tichy:2008du}. Other models have been developed by
Buonanno \etal~\cite{Buonanno:2007sv}, Lousto \etal \cite{Lousto:2009mf}, Barausse \& Rezzolla
\cite{Rezzolla:2007rz,Barausse:2009uz}, and others.

In the case of initially orbit-aligned spins, the initial parameter space is three-dimensional:
$\{q, S_1, S_2\}$. We use the simplest applicable formulas for the achieved end-state for an aligned-spin
system. The end-state mass formula we take from Eq.~(5) of \cite{Lousto:2009mf}:
\begin{align}
\Mf & = 1 - \eta E_{\rm ISCO} - E_2 \eta^2 - E_3 \eta^3  \nonumber \\
    &   - \frac{\eta^2}{(1+q)^2} \left[ E_S (\alpha_2+q^2 \alpha_1) + E_{\delta} (1-q) (\alpha_2 - q \alpha_1) \right. \nonumber \\
    &   \left. + E_A (\alpha_2 + q \alpha_1)^2 + E_D (\alpha_2 - q \alpha_1)^2 \right], \label{eq:mass_RIT}
\end{align}
where $\eta \equiv M_1 M_2/(M_1+M_2)^2 = q/(1+q)^2$ is the \emph{symmetric mass ratio} of the
binary, and the fitting parameters are:
\begin{align*}
E_{\rm ISCO} = & 1 - \frac{\sqrt{8}}{3} + 0.103803 \eta + \frac{(q (1 + 2 q) \alpha_1 + (2+q) \alpha_2)}{36 \sqrt{3} (1+q)^2} \\
               & + \frac{5 (q \alpha_1 - \alpha_2)^2}{162 \sqrt{2} (1+q)^2},\\
E_2          = & 0.341 , \;\; E_3 = 0.522 , \;\; E_S = 0.673 , \\
E_{\delta}   = & -0.36 , \;\; E_A = -0.014 , \;\; E_D = 0.26.
\end{align*}

For the final spin, one model with just enough complexity for our data sets here was given by
\cite{Rezzolla:2007rz,Barausse:2009uz} \footnote{Note that we have adapted Eq.~(4) of
\cite{Rezzolla:2007rz} to match our convention for $q$.}:
\beq
\alpf = \atil + s_4 \eta \atil^2 + s_5 \eta^2 \atil + t_0 \eta \atil + 2\sqrt{3} \eta + t_2 \eta^2 + t_3 \eta^3, \label{eq:spin_AEI} 
\eeq
where $\atil \equiv (q^2 \alpha_1 + \alpha_2)/(q^2+1)$ and the coefficients $\{s_4, s_5, t_0, t_2, t_3\}$ are:
\begin{align*}
s_4 &= -0.1229 \pm 0.0075,\; s_5 = 0.4537 \pm 0.1463,\nonumber\\
t_0 &= -2.8904 \pm 0.0359,\; t_2 = -3.5171 \pm 0.1210,\nonumber\\
t_3 &=  2.5763 \pm 0.4833.
\end{align*}

Using these formulae, we have constructed a set of configurations, which we present in
Table~\ref{tab:equiv_ID}, grouped by final Kerr spin.

\begin{table}\footnotesize
\caption{Final mass and spin of the post-merger Kerr BH, as measured by
radiation balance ($\Mf$, $\alpf$), and as predicted by phenomenological
equations \eqref{eq:mass_RIT}-\eqref{eq:spin_AEI}
($M_{\rm f, RIT}$,$\alpha_{\rm f, AEI}$). The final two columns give the
percentage relative error between the measured and predicted values, which
never exceeds 1.6 \% for the mass and 2.1 \% for the spin.}
\begin{tabular}{rrr|rr|rr}
\hline\hline
            run name &  $\Mf$ & $\alpf$& $M_{\rm f, RIT}$ & $\alpha_{\rm f, AEI}$ &   $\delta \Mf$ (\%) & $\delta \alpf$ (\%)\\
\hline
     \texttt{X1\_UU} & 0.9156 & 0.9053 & 0.9287 & 0.9112 &      1.43 &      0.65 \\
\hline
     \texttt{X1\_uu} & 0.9393 & 0.8119 & 0.9391 & 0.8038 &      0.03 &      0.99 \\
\hline
     \texttt{X1\_00} & 0.9520 & 0.6886 & 0.9497 & 0.6865 &      0.24 &      0.31 \\
     \texttt{X1\_UD} & 0.9505 & 0.6839 & 0.9359 & 0.6865 &      1.54 &      0.38 \\
\hline
   \texttt{X1.5\_00} & 0.9558 & 0.6664 & 0.9534 & 0.6644 &      0.25 &      0.30 \\
\hline
  \texttt{X1.75\_00} & 0.9588 & 0.6475 & 0.9565 & 0.6452 &      0.24 &      0.35 \\
\hline
     \texttt{X2\_00} & 0.9614 & 0.6254 & 0.9596 & 0.6244 &      0.19 &      0.17 \\
     \texttt{X2\_DU} & 0.9610 & 0.6120 & 0.9559 & 0.6244 &      0.54 &      2.02 \\
\hline
   \texttt{X2.5\_00} & 0.9671 & 0.5833 & 0.9654 & 0.5824 &      0.18 &      0.16 \\
\hline
     \texttt{X3\_00} & 0.9716 & 0.5432 & 0.9702 & 0.5429 &      0.15 &      0.07 \\
\hline
     \texttt{X4\_00} & 0.9782 & 0.4780 & 0.9812 & 0.4748 &      0.31 &      0.68 \\
     \texttt{X5\_U0} & 0.9816 & 0.4741 & 0.9773 & 0.4748 &      0.44 &      0.15 \\
     \texttt{X3\_d0} & 0.9737 & 0.4735 & 0.9720 & 0.4760 &      0.18 &      0.52 \\
     \texttt{X2\_D0} & 0.9683 & 0.4704 & 0.9649 & 0.4765 &      0.35 &      1.31 \\
     \texttt{X1\_DD} & 0.9646 & 0.4825 & 0.9674 & 0.4786 &      0.30 &      0.81 \\
\hline
     \texttt{X5\_00} & 0.9826 & 0.4186 & 0.9821 & 0.4202 &      0.06 &      0.37 \\
\hline
     \texttt{X6\_00} & 0.9857 & 0.3718 & 0.9854 & 0.3762 &      0.02 &      1.18 \\
     \texttt{X5\_D0} & 0.9834 & 0.3736 & 0.9791 & 0.3762 &      0.44 &      0.68 \\
     \texttt{X4\_D0} & 0.9803 & 0.3728 & 0.9791 & 0.3762 &      0.13 &      0.91 \\
     \texttt{X3\_D0} & 0.9762 & 0.3697 & 0.9739 & 0.3762 &      0.23 &      1.75 \\
     \texttt{X2\_DD} & 0.9718 & 0.3788 & 0.9729 & 0.3762 &      0.11 &      0.71 \\
\hline \hline
\end{tabular}  
\label{tab:final_params}
\end{table}

%\bibliographystyle{../bibtex/apsrev}
%
%\bibliography{../bibtex/references}

\end{document}